\documentclass[journal]{IEEEtran}

\usepackage{cite}
\usepackage{graphicx}
\usepackage{verbatim}
\usepackage{url}
\usepackage{amssymb}
\usepackage{mathrsfs}
\usepackage{amsmath}
\usepackage[sans]{dsfont}
\usepackage{stmaryrd}
\usepackage{rotating}
\usepackage{subfigure}
\usepackage[usenames]{color}
\usepackage{colortbl}

\definecolor{LightBlack}{rgb}{0.4,0.7,1}
\definecolor{BricRed}{rgb}{0,0,0}
\definecolor{OlGreen}{rgb}{0.216,0.6,0.274} 

\newtheorem{lemma}{Lemma}
\newtheorem{theorem}{Theorem}
\newtheorem{corollary}[lemma]{Corollary}

\ifodd 1
\newcommand{\com}[1]{\textbf{\color{blue} (COMMENT: #1)}} 
\else

\newcommand{\com}[1]{}
\fi

\begin{document}

\title{\mbox{Capacity of Large-scale CSMA Wireless Networks}}

\author{Chi-Kin~Chau, Minghua~Chen, and Soung Chang Liew
\thanks{
C.-K. Chau was supported by a Croucher Foundation fellowship. M. Chen and S. C. Liew are supported by Competitive Earmarked Research Grants (Project \# 411008, 411209, and 414507) established under the University Grant Committee of the Hong Kong SAR, China. M. Chen is also supported by a Direct Grant (Project \# 2050397) of The Chinese University of Hong Kong (CUHK). Major part of the work was done when C.-K. Chau visited CUHK, under the support of the Direct Grant (Project \# 2050397) of CUHK. A preliminary version was presented at ACM MobiCom 09'.}
\thanks{C.-K. Chau is with Computer Laboratory, University of Cambridge, UK (e-mail: chi-kin.chau@cl.cam.ac.uk).}
\thanks{M. Chen and S.-C. Liew are with Department of Information Engineering at the Chinese University of Hong Kong, Shatin, Hong Kong, China (e-mail: \{minghua,soung\}@ie.cuhk.edu.hk).}
}

\maketitle

\begin{abstract}
In the literature, asymptotic studies of multi-hop wireless network capacity often consider only centralized and deterministic TDMA (time-division multi-access) coordination schemes. There have been fewer studies of the asymptotic capacity of large-scale wireless networks based on CSMA (carrier-sensing multi-access), which schedules transmissions in a distributed and random manner. With the rapid and widespread adoption of CSMA technology, a critical question is that whether CSMA networks can be as scalable as TDMA networks. To answer this question and explore the capacity of CSMA networks, we first formulate the models of CSMA protocols to take into account the unique CSMA characteristics not captured by existing interference models in the literature. These CSMA models determine the feasible states, and consequently the capacity of CSMA networks. We then study the throughput efficiency of CSMA scheduling as compared to TDMA. Finally, we tune the CSMA parameters so as to maximize the throughput to the optimal order. As a result, we show that CSMA can achieve throughput as $\Omega(\frac{1}{\sqrt{n}})$, the same order as optimal centralized TDMA, on uniform random networks. Our CSMA scheme makes use of an efficient backbone-peripheral routing scheme and a careful design of dual carrier-sensing and dual channel scheme. We also address the implementation issues of our CSMA scheme. 
\end{abstract}


\section{Introduction}

An important characteristic that distinguishes wireless networks from wired networks is the presence of {\em spatial interference}, wherein the transmission between a pair of nodes can upset other transmissions in its neighborhood. Spatial interference imposes a limit on the capacity of wireless networks.

The seminal paper \cite{GK00capacity} by Gupta and Kumar revealed that the capacity of wireless networks constrained by spatial interference is upper bounded by ${\rm O}\big(\frac{1}{\sqrt{n}}\big)$ for $n$ number of mutually communicating nodes on a uniform random network, regardless of the chosen scheduling and routing schemes. Many similar upper bounds are derived for more sophisticated settings (e.g., with optimal source and network coding schemes \cite{XK06scale}). In~\cite{dai08throughput}, Dai and Lee derived the upper bound ${\rm O}\big(\frac{1}{\sqrt{n}}\big)$ for multi-hop random access networks using a simple queuing analytical argument. They also showed that this upper bound is achievable only if the maximum throughput of each  local node is a constant independent of $n$.

Since then, a number of solutions have been proposed to achieve the upper bounds in various settings. Particularly, \cite{FDTT00percolation} showed that by an efficient backbone-peripheral routing scheme (analogously called ``highway system'') and a two-stage TDMA scheme, ${\Omega}\big(\frac{1}{\sqrt{n}}\big)$ is achievable on a uniform random network with high probability.

So far, the studies of achievable wireless capacity in the literature consider only centralized controls and a-priori scheduling schemes with TDMA. On the practical front, carrier-sensing multi-access (CSMA) networks (e.g., Wi-Fi), which make use of distributed and randomized medium-access protocols, are receiving wide adoption across enterprises and homes. It is not clear whether the results related to centrally-scheduled networks are directly applicable to CSMA networks.

To bridge the gap between practice and research, it will be interesting to find out to what extent the capacity of CSMA networks can be scaled. In particular, can the simple distributed scheduling of CSMA scales network capacity as well as central scheduling can?

The answer, according to our study, is ``yes''. However, the way to go about achieving CSMA scalability is non-trivial and several mechanisms must be in place before scalability can be attained. For example, the use of dual carrier-sensing ranges in two channels will be needed; and one must be able to assign different back-off countdown times to different nodes in a distributed manner.

To establish our results, besides building on the past work of others, we find it necessary to clarify and add rigor to the previous frameworks. It is well known that spatial interference imposes a constraint on the links that can be active simultaneously. Given an interference model, in general there can be a number of subsets of links that can be active simultaneously. Each such subset of links is called a {\em feasible state}. For a central scheduler, all feasible states are available for the design of its schedule\footnote{A schedule is a sequence of feasible states that are active at different times.}. For CSMA networks, its distributed nature does not allow us to dictate which particular feasible state will be active at what time. The problem becomes even more challenging because if not designed properly, CSMA may allow a subset of links that is not interference-safe to transmit simultaneously, leading to the so-called {\em hidden-node problem}.

We define the feasible states allowed by the CSMA protocol in a rigorous manner. We argue that the hidden-node problem in CSMA networks is caused by a mismatch between the feasible states allowed by CSMA and the feasible states of an underlying interference model. We show how to resolve this mismatch to create hidden-node free CSMA networks. Most importantly, we show that hidden-node free CSMA networks can achieve the same scaling of throughput as the central scheduler provided the aforementioned dual carrier-sensing and dual channel scheme is in place. Our capacity-optimal CSMA scheme not only demonstrates the theoretical achievable throughput of CSMA networks, but also outlines a practical way to achieve it.



\section{Background and Overview}

The basic idea of CSMA is that before a transmitter attempts its transmission, it needs to infer the channel condition by sensing the channel. If it infers that its transmission will upset (or be upset by) any receiver's on-going transmissions (including its own receiver), then it defers its transmission. In addition, to prevent two transmitters from beginning their transmissions at the same time (given that they both sense the channel to be safe for transmission), each transmitter undergoes a random backoff count-down period before transmission. The count-down will be frozen when channel is sensed to be not interference-safe (i.e., transmission is collision-prone), and will be resumed when the channel is sensed to be interference-safe again. A transmission will be considered successful, when the transmitter can receive an ACK packet by the corresponding receiver, upon the completion of transmission.

Compared to the centralized TDMA scheme, the CSMA  protocol has two distinguishing characteristics:
\begin{enumerate}

\item[i)]  CSMA is an ACK-based protocol, in which the receivers are required to reply an ACK packet for each successful transmission. Thus, {\em bi-directional communications} need to be explicitly considered when formulating the constraints on simultaneous transmission imposed by CSMA. The centralized TDMA schemes in prior work \cite{GK00capacity,XK06scale,FDTT00percolation,OLT07mimo,JLNP07scaling,FMM07limit}, however, did not consider bi-directional communications and ACK packets.

\item[ii)] CSMA is a distributed random access protocol. Each transmitter chooses a random time instance to initiate its transmission, and it can only rely on its limited
local knowledge to infer whether its transmission is compatible with other simultaneous transmission under various interference settings\footnote{Note that the interference is not necessarily symmetric --- a transmission could upset another simultaneous transmission but not the converse.}.
Unlike the centralized TDMA schemes, such a distributed control requires only limited a-priori coordination among
transmitters and receivers.


\end{enumerate}



Despite the popularity of CSMA protocols, capacity analysis applicable to large-scale CSMA wireless networks receive relatively little attention in the literature. A likely reason could be that CSMA protocols are generally regarded as synonymous to the so-called ``protocol model'' in many TDMA based papers. The ``protocol model'' is, in fact, a simplified pairwise interference model that serves to model interference among simultaneous links, which neither explicitly considers nor precisely models the aforementioned characteristics i)-ii) of CSMA\footnote{Gupta and Kumar's seminal paper \cite{GK00capacity} appears to be the first to coin the phrase ``protocol model'', but without specifying any distributed protocol that can implement the protocol model, other than centralized schemes by TDMA.}. As such, it is not clear 1) whether the capacity results based on these interference models can apply to CSMA networks; and 2) whether CSMA can achieve the same throughput performance as centralized TDMA.

There is a considerably large body of literature about single-hop CSMA networks \cite{JL08hidden,LKLW09boe}. Here we study the more general multi-hop CSMA networks, the results of which are quite limited in the literature~\cite{GCL06WirelessMultiHopThput,NL07ThputMultiHop}. We also note that \cite{BBM06Aloha} has studied the capacity of multi-hop Aloha networks. However, Aloha protocol is different from CSMA protocol as it has no carrier-sensing operations. Also, the definition of capacity in \cite{BBM06Aloha} appears to be different from the conventional Gupta-Kumar's one \cite{GK00capacity,XK06scale,FDTT00percolation,LLL08multicast,JLNP07scaling,SCB10greenwave}.

\begin{table*}[hbt] 
  \centering
\begin{tabular}{c|c|c|c|c|c|c}
\hline
 & \multicolumn{2}{c|}{Uni-directional feasible family} & \multicolumn{2}{c|}{Bi-directional feasible family} & \multicolumn{2}{c}{Carrier-sensing feasible family}\tabularnewline
\hline
 & Pairwise  & Aggregate  & Pairwise  & Aggregate  & \multicolumn{2}{c}{Pairwise} \tabularnewline
  & interference  & interference  & interference  & interference  & \multicolumn{2}{c}{carrier-sensing} \tabularnewline
\hline
\hline
Random   & \multicolumn{2}{c|}{Upper bound: $O(\frac{1}{\sqrt{n}})$~\cite{GK00capacity}} &  \multicolumn{2}{c|}{Upper bound: $O(\frac{1}{\sqrt{n}})$ (this paper)}  & \multicolumn{2}{c}{Upper bound: $O(\frac{1}{\sqrt{n}})$ (this paper)} \tabularnewline
network  & \multicolumn{2}{c|}{Achievable as: $\Omega(\frac{1}{\sqrt{n}})$ }  &  \multicolumn{2}{c|}{Achievable as: $\Omega(\frac{1}{\sqrt{n}})$}  &  \multicolumn{2}{c}{Achievable as: $\Omega(\frac{1}{\sqrt{n}})$} \tabularnewline
capacity &  \multicolumn{2}{c|}{by TDMA~\cite{FDTT00percolation}}  &  \multicolumn{2}{c|}{by TDMA (this paper)}
  &  \multicolumn{2}{c}{ by dual carrier sensing (this paper)}  \tabularnewline
\hline
\end{tabular} 
  \caption{Capacity of uniform random networks over various feasible families.}\label{Tab:fea.family.and.capacity}
\end{table*}

\subsection{Outline of Our Results}

To explore the capacity of CSMA networks, we first formulate the models of CSMA
protocols to take into account characteristics i)-ii). These models determine the upper and lower bound on the capacity of CSMA networks, and are functions of various CSMA parameters. We then study the throughput efficiency of CSMA relative to TDMA, following the same procedure as in~\cite{LKLW09boe} and~\cite{JW08csma}. Finally we tune the CSMA parameters so that the capacity of a CSMA network is maximized to the optimal order. Our approach is divided into four parts:

1) {\em Formulation of Carrier-sensing Decision Model} (Sec.~\ref{sec:models}):
Our models for CSMA protocol consist of two components that capture two major functionalities of CSMA.

\begin{itemize}
  \item The {\em decision model} that
formally formulates the constraints on simultaneously active links imposed by CSMA carrier sensing operations, such as distance-based carrier sensing. We explicitly distinguish the decision model of CSMA protocols from the interference model. For instance, the fact that two simultaneously active links are allowed by CSMA does not necessarily mean that they do not interfere with each other. This is the well-known hidden node problem.


  \item The {\em random access scheme} that captures
how CSMA access the wireless air time and space. The key challenge is to understand the throughput efficiency of distributed and randomized channel access mechanism of CSMA, as compared to centralized TDMA scheme.
\end{itemize}

We establish the relationship between our CSMA models and the interference models from the literature in Sec.~\ref{sec:models}. 

2) {\em Hidden-node-free Design of CSMA Networks} (Sec.~\ref{sec:hidden}): There are various interference models in the literature (including the so-called ``protocol model''). They are intended to capture uni-directional transmissions where ACK packets are not required. In this paper, we extend these interference models to the setting of bi-directional transmissions, under which CSMA protocols typically operate.

It is well-known that the distributed transmission scheduling in CSMA may not be able to prevent spatial interference, as known as the {\em hidden node problem} \cite{JL08hidden}. Utilizing our proposed carrier-sensing decision models, we formally define the hidden node problem as due to a carrier-sensing decision model violating the feasibility of a bi-directional interference model. Furthermore, we derive sufficient conditions for pairwise carrier-sensing decision model (based on carrier-sensing range) to eliminate the hidden node problem under various interference settings (Theorem~\ref{thm:hidden}). Our results include the prior one in \cite{JL08hidden} as a special case.

By eliminating the hidden-node problem, we can apply elegant mathematical tools to analyze the capacity and throughput performance of multi-hop CSMA networks.

3) {\em Stationary State Analysis of Random Access} (Sec.~\ref{sec:random}): To study the behavior of the random access scheme, we consider an idealized version of IEEE 802.11 DCF based on a continuous-time Markov chain model in order to capture the essential features of CSMA. This continuous Markov chain model has been used in various analyses \cite{LKLW09boe,JW08csma,WK05throughput}. 

Based on the hidden-node-free design of CSMA networks, the long-term throughput of CSMA with random access is characterized by the stationary distribution of the continuous-time Markov chain model. Following the same procedure as in \cite{LKLW09boe,JW08csma,WK05throughput}, we present the stationary distribution, and hence, the long-term throughput of hidden-node-free CSMA networks under various carrier-sensing decision models in Sec.~\ref{sec:random}. We also show that CSMA random access schemes can be tuned to perform as well as TDMA schemes.


4) {\em Design of Dual Carrier-Sensing} (Secs.~\ref{sec:net}-\ref{sec:multi}):
On hidden-node-free CSMA networks, we show that the current CSMA setting with a single homogeneous carrier-sensing operation fails to achieve the optimal capacity ${\Omega}(\frac{1}{\sqrt{n}})$ on a uniform random network. It can at most achieve a capacity of ${\rm O}(\frac{1}{\sqrt{n \log n}})$ with high probability (shown by Theorem~\ref{thm:bd_single}).

We then show that the design of dual carrier-sensing operations can achieve the capacity of the same order as optimal centralized TDMA. Our design is drawn from an efficient backbone-peripheral routing scheme in \cite{FDTT00percolation}, based on which we show that using two different carrier-sensing ranges are sufficient to achieve optimal capacity of ${\Omega}(\frac{1}{\sqrt{n}})$ on a uniform random network with high probability (shown by Theorem~\ref{thm:bd_dual}).

In this paper, we not only provide insights for the optimal asymptotic capacity of wireless networks by our dual carrier-sensing scheme, but also address practical issues of implementing our scheme. First, we address the scalability issue during the dynamic switching between the dual carrier-sensing operations. We propose to use two frequency channels to distinguish the two carrier-sensing operations. Second, we address the issue of half-duplexity across two frequency channels, which enables low-cost implementation of our scheme.

We summarize our results and related work in Table~\ref{Tab:fea.family.and.capacity}.

\section{Formulation and Models} \label{sec:models}

First, note that some key notations are listed in Table~\ref{tab1}.

\begin{table}[hbt] 
\caption{Key Notations}\label{tab1} 
\centering
\begin{tabular}
[l]{l|l}
\hline
\textbf{Notation} &  \textbf{Definition}\\
\hline
$N^{\sf sd}$ & Set of source-sink pairs.\\
$\lambda_k$ & Data rate of source-sink pair $k\in N^{\sf sd}$.\\
${X}$ & Set of relaying links induced by the paths\\
    &  between all source-sink pairs in $N^{\sf sd}$.\\
$t_i$ & Coordinates of the transmitter of link $i\in X$.\\
$r_i$ & Coordinates of the receiver of link $i\in X$.\\
${\cal S}$ & {\em Feasible state}, a subset of links that \\
    & can simultaneously transmit. \\
${\mathscr F}, {\mathscr U}, {\mathscr B}, {\mathscr C}$ & {\em Feasible family}, a set of feasible states. \\
\hline
${\sf P}_{\sf tx}$ & Fixed transmission power of all nodes. \\
${\sf N}_{0}$ & Fixed noise power. \\
$\alpha$ & Power decaying factor in radio transmission.\\
$\beta$ & Minimum Signal-to-Interference-Noise ratio  \\
    &   for successful receptions.\\
$\Delta$ & Guard-zone coefficient, used in noise-absence \\
    & pairwise SIR interference model. \\
$ {\sf r}_{\sf xcl}$ & Interference range, used in fixed range \\
    & interference models.\\
$ {\sf r}_{\sf tx}$ &Communication range, used in fixed range \\
    & interference models. \\
$ {\sf r}_{\sf cs}$ & Carrier sensing range, used in pairwise \\
    & CSMA decision models. \\
$ {\sf t}_{\sf cs}$ & Carrier sensing power threshold, used in  \\
    & aggregate CSMA decision models. \\
\hline
\end{tabular} 
\end{table}

A central problem of multi-hop wireless communications is defined as follows. Given a set of source-sink pairs $N^{\sf sd}$ and a set of data rate $\{\lambda_k, k\in N^{\sf sd}\}$, we ask whether successful wireless communications can be established between all the sources and sinks in $N^{\sf sd}$ to sustain the required rate $\{\lambda_k, k\in N^{\sf sd}\}$, possibly using other nodes as relays, subject to a certain interference model of simultaneous wireless transmissions.

Specifically, we consider the following two degrees of freedom
in establishing the wireless communications:
\begin{enumerate}

\item {\em Routing scheme} that selects the appropriate relaying nodes to connect the sources and sinks.

\item {\em Scheduling scheme} that assigns (deterministically or randomly) the opportunities of transmissions at relaying nodes.

\end{enumerate}
Furthermore, these wireless communications should be established in a distributed manner with minimal global knowledge and coordination among the nodes.

Hence, we first present several common interference models of feasible simultaneous wireless transmissions. Then we extend these interference models to the setting of bi-directional communications. Next, we formulate carrier-sensing decision models that capture distributed control of transmissions.

\subsection{Interference Models}



An {\em interference model} is defined by its interference-safe feasible family. Some common interference-safe feasibility families in the literature are defined as follows. To simplify the definitions, we implicitly assume $i \ne j$.
\begin{enumerate}

\item[${\sf a.0}$)] {\em Pairwise fixed-range feasible family}:

${\cal S} \in {\mathscr U}^{\sf pw}_{\rm fr}\big[{X},{\sf r}_{\sf xcl}, {\sf r}_{\sf tx}\big]$, if and only if  for all $i, j \in {\cal S}$,
\begin{equation}
|t_j - r_i| \ge {\sf r}_{\sf xcl} \mbox{ \ and \ } |t_i - r_i| \le {\sf r}_{\sf tx}
\end{equation}

\item[${\sf a.1}$)] {\em Pairwise (noise-absent) SIR feasible family}:

${\cal S} \in {\mathscr U}^{\sf pw}_{\rm sir}\big[{X},\Delta\big]$, if and only if  for all $i, j \in {\cal S}$,
\begin{equation}
|t_j - r_i| \ge (1 + \Delta) |t_i - r_i|
\end{equation}

\item[${\sf a.2}$)] {\em Pairwise SINR feasible family}:

${\cal S} \in {\mathscr U}^{\sf pw}_{\rm sinr}\big[{X},\beta\big]$, if and only~if for all $i, j \in {\cal S}$,
\begin{equation}
 \frac{{\sf P}_{\sf tx} |t_i - r_i|^{-\alpha}}{{\sf N_0} +
 {\sf P}_{\sf tx}  |t_j - r_i|^{-\alpha}} \ge \beta
\end{equation}

\item[${\sf a.3}$)] {\em Aggregate SINR feasible family}:

${\cal S} \in {\mathscr U}^{\sf ag}_{\rm sinr}\big[{X},\beta\big]$, if and only if  for all $i \in {\cal S}$,
\begin{equation}
 \frac{{\sf P}_{\sf tx} |t_i - r_i|^{-\alpha}}{{\sf N_0} + \underset{j \in {\cal S} \backslash \{ i \}}{\sum} {\sf P}_{\sf tx}  |t_j - r_i|^{-\alpha}} \ge \beta
\end{equation}

\end{enumerate}
Also, we suppose ${\sf r}_{\sf xcl} > {\sf r}_{\sf tx}$, $\Delta > 0$, $\alpha > 2$, $\beta > 0$, and uniform power ${\sf P}_{\sf tx}$ at all nodes. For ${\sf a.2}$)-${\sf a.3}$), ${\sf P}_{\sf tx}|t_i - r_i|^{-\alpha} \ge \beta {\sf N_0}$ for all $i \in {X}$. Otherwise, $t_i$ cannot successfully transmit packets to $r_i$ even without interference.

The notion of feasible family applies to both pairwise and aggregate interference models. Pairwise models ${\sf a.0}$)-${\sf a.2}$) can be captured by the use of conflict graph,
whereas the notion of feasible family is more generally applicable to ${\sf a.0}$)-${\sf a.3}$).


In \cite{GK00capacity}, pairwise SIR interference model ${\sf a.1}$) is called ``protocol model'', whereas aggregate SINR interference model ${\sf a.3}$) is called ``physical model''. The naming in this paper emphasizes the interference of transmissions, and avoids confusion with CSMA protocol models\footnote{We remark that \cite{AK04bound} also presents a ``generalized protocol model'' with arbitrary interference footprint around the transmitters that models more general pairwise interference settings, and a ``generalized physical model'' that specifically applies to the Gaussian channel.}.

\subsection{Bi-directional Interference Models}

\begin{figure}[htb!] 
    \centering
    \includegraphics[scale=0.5]{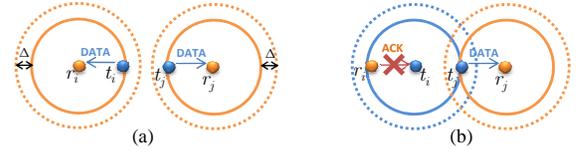} 
  \caption{In Fig. (a) the normal DATA packet transmissions from transmitters will not interfere with each other, but in Fig. (b) there is interference when transmitting ACK packet.} \label{fig:uni-bi} 
\end{figure}

The interference-safe constraints ${\sf a.0}$)-${\sf a.3}$) are uni-directional, based on the assumption that the receiver is not required to reply an ACK packet to the transmitter upon a successful transmission. For ACK-based transmissions, interference can occur between two transmitters, between two receivers, and between a transmitter and a receiver. See Fig.~\ref{fig:uni-bi} for an example of pairwise SIR interference model. Without the reception of ACK packets, the transmitter will consider the transmission unsuccessful and retransmit the DATA packet later on. Hence, we need to ensure that the transmissions of DATA packets and ACK packets of all simultaneous links do not interfere with each other.

Let ${\sf dist}(i,j) \triangleq \min(|t_j - r_i|, |r_j - t_i|, |r_j - r_i|, |t_j - t_i|)$. We consider the bi-directional versions of interference-safe constraints as follows.

\begin{enumerate}

\item[${\sf b.0}$)] {\em Bi-directional pairwise fixed-range feasible family}:

${\cal S} \in {\mathscr B}^{\sf pw}_{\rm fr}\big[{X},{\sf r}_{\sf xcl}, {\sf r}_{\sf tx}\big]$, if and only if  for all $i, j \in {\cal S}$,
\begin{equation}
{\sf dist}(i,j) \ge {\sf r}_{\sf xcl} \mbox{ \ and \ } |t_i - r_i| \le {\sf r}_{\sf tx}
\end{equation}

\item[${\sf b.1}$)] {\em Bi-directional pairwise SIR feasible family}:

${\cal S} \in {\mathscr B}^{\sf pw}_{\rm sir}\big[{X},\Delta\big]$, if and only if for all $i, j \in {\cal S}$,
\begin{equation}
{\sf dist}(i,j) \ge (1 + \Delta) |t_i - r_i|
\end{equation}

\item[${\sf b.2}$)] {\em Bi-directional pairwise SINR feasible family}:

${\cal S} \in {\mathscr B}^{\sf pw}_{\rm sinr}\big[{X},\beta\big]$, if and only if for all $i, j \in {\cal S}$,
\begin{equation}
 \frac{{\sf P}_{\sf tx} |t_i - r_i|^{-\alpha}}{{\sf N_0} +
 {\sf P}_{\sf tx}  \big( {\sf dist}(i,j) \big)^{-\alpha}} \ge \beta
\end{equation}

\item[${\sf b.3}$)] {\em Bi-directional aggregate SINR feasible family}
\footnote{A more precise definition should replace the denominator of the LHS of Eqn.~(\ref{eqn:b.3}) by
$
{\sf N_0} \mbox{+} \min \big\{
{\sum}_{j \in {\cal S} \backslash \{ i \}} {\sf P}_{\sf tx} \big( \min\{|t_j \mbox{-} t_i|, |r_j \mbox{-} t_i| \} \big)^{-\alpha},$
${\sum}_{j \in {\cal S} \backslash \{ i \}} {\sf P}_{\sf tx} \big( \min\{|t_j \mbox{-} r_i|, |r_j \mbox{-} r_i| \} \big)^{-\alpha}\big\}.
$
Here we choose the simpler and more conservative form in Eqn.~(\ref{eqn:b.3}), as it is sufficient for our results.}:

${\cal S} \in {\mathscr B}^{\sf ag}_{\rm sinr}\big[{X},\beta\big]$, if and only if for all $i \in {\cal S}$,
\begin{equation}\label{eqn:b.3}
 \frac{{\sf P}_{\sf tx} |t_i - r_i|^{-\alpha}}{{\sf N_0} + \underset{j \in {\cal S} \backslash \{ i \}}{\sum} {\sf P}_{\sf tx} \big( {\sf dist}(i,j) \big)^{-\alpha}} \ge \beta
 \end{equation}

\end{enumerate}

Compared with the uni-direction interference-safe constraints, the bi-directional counterparts consider the interference from both the DATA and ACK transmissions.

\subsection{Carrier-Sensing Decision Models}

The interference-safe constraints ${\sf a.0}$)-${\sf a.3}$) and ${\sf b.0}$)-${\sf b.3}$) capture the global spatial interference in the network. In CSMA, a transmitter has only local knowledge of its interference condition, but not the interference conditions at its targeted receiver or at the transmitting and receiving nodes of other active links. The decision of a transmitter whether to transmit is only determined by its carrier-sensing operation, rather than by the global knowledge of spatial interference.

We define {\em carrier-sensing decision models}, in which a feasible family is a set of links that may transmit simultaneously under a carrier sensing operation. But this feasible family may or may not be interference-safe under the uni-/bi-directional interference models. We present two feasible families to capture carrier-sensing operations, defined as follows.
\begin{enumerate}

\item[${\sf c.1}$)] {\em Pairwise carrier-sensing feasible family}:

${\cal S} \in {\mathscr C}^{\sf pw}\big[{X},{\sf r}_{\rm cs}\big]$, if and only~if for all $i, j \in {\cal S}$,
\begin{equation}
|t_j - t_i| \ge {\sf r}_{\rm cs}
\end{equation}

\end{enumerate}
{\color{Black} In pairwise carrier-sensing decision model ${\sf c.1}$), transmission decision is based on the distance from other simultaneous transmitters. ${\sf c.1}$) is often used together with the pairwise interference model for analysis in the literature. In fact, in analysis and in actual implementation, ${\sf c.1}$) is also compatible with the aggregate interference model.
For instance, \cite{FLH10IPCS} recently introduced a novel and practical approach to implement pairwise
carrier-sensing decision model ${\sf c.1}$) with respect to aggregate interference model, using
Incremental-Power Carrier Sensing (IPCS). 

The basic idea of {\em Incremental-Power Carrier-Sensing (IPCS)} is that a transmitter $i$ can estimate the distance to an individual simultaneously active transmitter $k$ by measuring the change of interference level. Suppose that initially $i$ measures the aggregate interference level as:
$\big( {\sf N_0} + \sum_{j \in {\cal S}\backslash \{ k\}} {\sf P}_{\sf tx} |t_j - t_i|^{-\alpha} \big)$. Then when $k$ transmits, the measured change of interference level at $i$ becomes $\Delta{\sf P}_i = {\sf P}_{\sf tx} |t_k - t_i|^{-\alpha}$, which reveals the distance to $k$. This mechanism proceeds as follows. Hence, each transmitter $i$ requires to maintain a counter ${\sf cnt}_i$ (initially set as 0). When $i$ detects any change $\Delta{\sf P}_i$,
\begin{itemize}
\item if $\Delta{\sf P}_i \ge {\sf P}_{\sf tx} {\sf r}_{\rm cs}^{-\alpha}$,
then ${\sf cnt}_i \leftarrow {\sf cnt}_i + 1$.
\item if $\Delta{\sf P}_i \le -{\sf P}_{\sf tx} {\sf r}_{\rm cs}^{-\alpha}$,
then ${\sf cnt}_i \leftarrow {\sf cnt}_i - 1$.
\end{itemize}
Transmitter $i$ is allowed to transmit only if ${\sf cnt}_i = 0$. Suppose that there is no transmitters that will simultaneously start to transmit at the same time\footnote{This will be true, when we use continuous exponentially random count-down as in the next section.}, IPCS can realize pairwise carrier-sensing decision model ${\sf c.1}$).

}

{\color{Black}

On the other hand, the current IEEE 802.11 networks uses a power-threshold based carrier sensing mechanism, such that a transmitter decides its transmissions based on the aggregate interference level measured before the transmission:
\begin{enumerate}

\item[${\sf c.2}$)] {\em Aggregate carrier-sensing feasible family}:

${\cal S} \in {\mathscr C}^{\sf ag}\big[{X},{\sf t}_{\rm cs}\big]$, if and only if there is a sequence $(i_1 , ..., i_{|{\cal S}|})$, such that for each $i_k \in {\cal S}$
\begin{equation}
{\sf N_0} + \underset{j \in \{i_1,..., i_{k-1}\}}{\sum} {\sf P}_{\sf tx}
|t_j - t_{i_k}|^{-\alpha} \le {\sf t}_{\rm cs}
\end{equation}

\end{enumerate}
That is, when each transmitter $i_k$ sees the aggregate interference level from other simultaneously active transmitters that have started transmission before is below the power threshold ${\sf t}_{\rm cs}$, $i_k$ decides that it is allowed to transmit.

Although aggregate carrier-sensing decision model ${\sf c.2}$) is easier to implement than pairwise carrier-sensing decision model ${\sf c.1}$) (which relies on IPCS), pairwise carrier-sensing does not depend on the order of decision sequence of transmitters, which is more amenable to analysis.
}

\section{Hidden-node-free Design} \label{sec:hidden}

Using only local interference conditions, the local decisions of transmissions in CSMA cannot completely prevent harmful spatial interference (i.e., the hidden-node problem), or may sometimes over-react to benign spatial interference (i.e., the exposed-node problem). While they are well recognized in the literature, lacking are formal definitions that comprehensively consider various interference and carrier-sensing decision models. Here, we provide formal definitions to hidden-node and exposed-node problems based on the models in Sec.\ref{sec:models}. We then also provide sufficient conditions to eliminate the hidden-node problem.

Because CSMA is an ACK-based protocol, we consider a bi-directional interference-safe feasible family ${\mathscr B}\big[X\big] $ from one of ${\sf b.0}$)-${\sf b.3}$). Given a carrier-sensing feasible family ${\mathscr C}\big[X\big]$ from one of ${\sf c.1}$)-${\sf c.2}$), we define
\begin{itemize}

\item {\em Hidden-node problem}: if ${\mathscr B}\big[X\big] \not\supseteq {\mathscr C}\big[X\big]$

\item {\em Exposed-node problem}: if ${\mathscr C}\big[X\big]  \not\supseteq {\mathscr B}\big[X\big]$

\end{itemize}

\begin{figure}[htb!]
    \centering
      \includegraphics[scale=0.5]{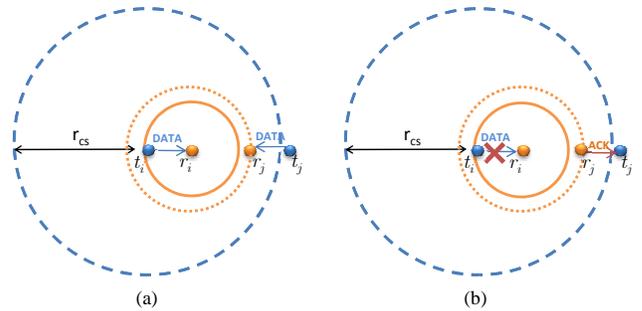}
  \caption{In Fig. (a) the carrier-sensing decision model correctly permits the simultaneous links for DATA packet transmission, but fails in the case of ACK packet transmission in Fig. (b). Hence, ${\mathscr B}^{\sf pw}_{\rm sir}\big[X\big] \not\supseteq {\mathscr C}^{\sf pw}\big[X\big]$} \label{fig:cs}
\end{figure}

Namely, hidden-node problem refers to situations where the carrier-sensing decision violates the bi-directional interference-safe constraints, whereas exposed-node problem is where the carrier-sensing decision is overly conservative in attempting to conform to the bi-directional interference-safe constraints. Our definitions naturally generalize the ones in \cite{JL08hidden}, which considers only pairwise interference and carrier-sensing decision models. For example, we illustrate an instance of hidden-node problem for pairwise carrier-sensing decision model and pairwise SIR interference model in Fig.~\ref{fig:cs}.

As studied in \cite{JL08hidden, WK05throughput}, hidden-node problem causes unfairness in CSMA networks. In this paper, we only consider CSMA networks that are designed to be hidden-node free. Besides the benefit of better fairness, more importantly, the overall performance of a hidden-node free CSMA network is tractable analytically. For example, the crucial Eqn.~(\ref{eq:stationary.dist}) of CSMA stationary states to be presented in Sec.~\ref{sec:random} is valid only for a CSMA network that is hidden-node free.

One of our contributions is to establish the sufficient conditions to eliminate hidden-node problem in various interference models. {\color{Black} We note that it is more complicated to design aggregate carrier-sensing decision model ${\sf c.2}$) to prevent hidden nodes. Hence, in the following we only consider pairwise carrier-sensing decision model ${\sf c.1}$).}
\\


\begin{lemma} \label{lem:pw_ag}
If $\Delta \le \beta^{\frac{1}{\alpha}}-1$, then
\begin{equation}
{\mathscr U}^{\sf pw}_{\rm sir}[X,\Delta] \supseteq {\mathscr U}^{\sf pw}_{\rm sinr}[X,\beta]
\supseteq {\mathscr U}^{\sf ag}_{\rm sinr}[X,\beta]
\end{equation}
\end{lemma}

\begin{lemma} \label{lem:ag_fr}
Let ${\sf r}_{\sf tx} = {\max}_{i \in X} |t_i - r_i|$. If
\begin{equation}
\begin{array}{c}
{\sf r}_{\sf xcl} \ge
\Big(  \frac{1}{{\sf P}_{\sf tx} {\sf k}(\alpha)} \big( \frac{{\sf P}_{\sf tx}}{\beta}{\sf r}_{\sf tx}^{-\alpha} - {\sf N_0} \big) \Big)^{-\frac{1}{\alpha}} + {\sf r}_{\sf tx}
\end{array}
\end{equation}
where ${\sf k}(\alpha) \triangleq \sum_{k=1}^{\infty} 4 \lceil \pi (2k + 2) \rceil k^{-\alpha}$,
then
\begin{equation}
{\mathscr U}^{\sf ag}_{\rm sinr}[X,\beta] \supseteq  {\mathscr U}^{\sf pw}_{\rm fr}\big[{X},{\sf r}_{\sf xcl}, {\sf r}_{\sf tx}\big]
\end{equation}
\end{lemma}

\begin{figure}[htb!] 
\centering
\mbox {
    \subfigure[\small{}]{\includegraphics[width=110pt]{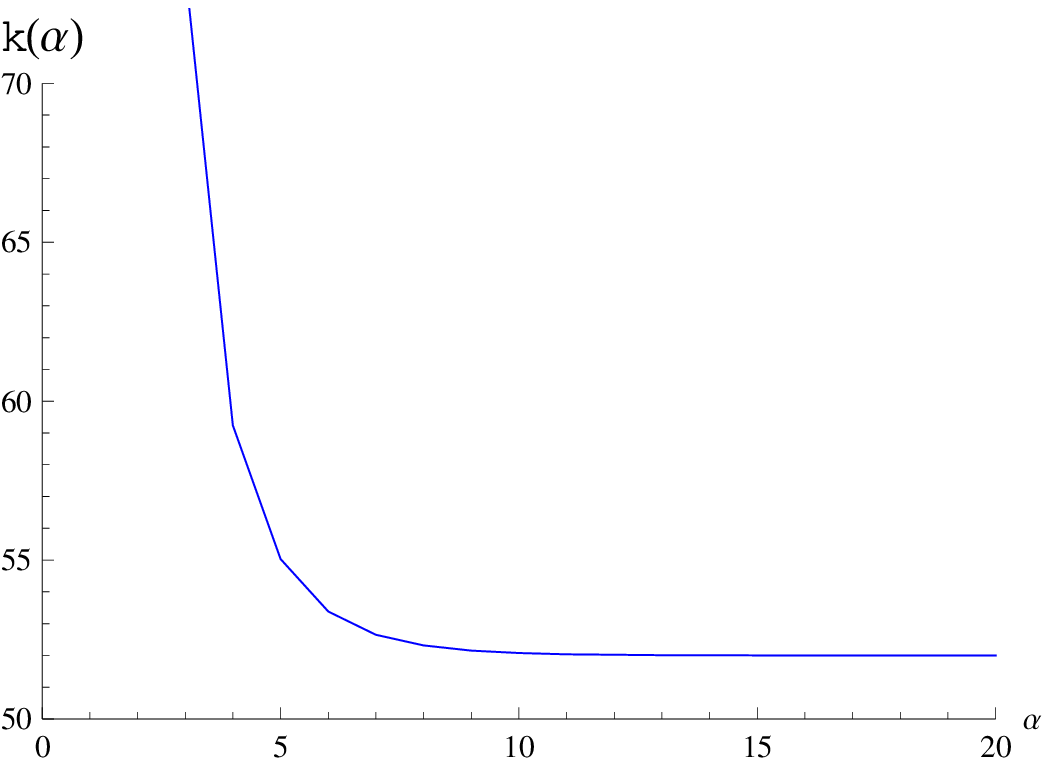}}
    \subfigure[\small{}]{\includegraphics[width=110pt]{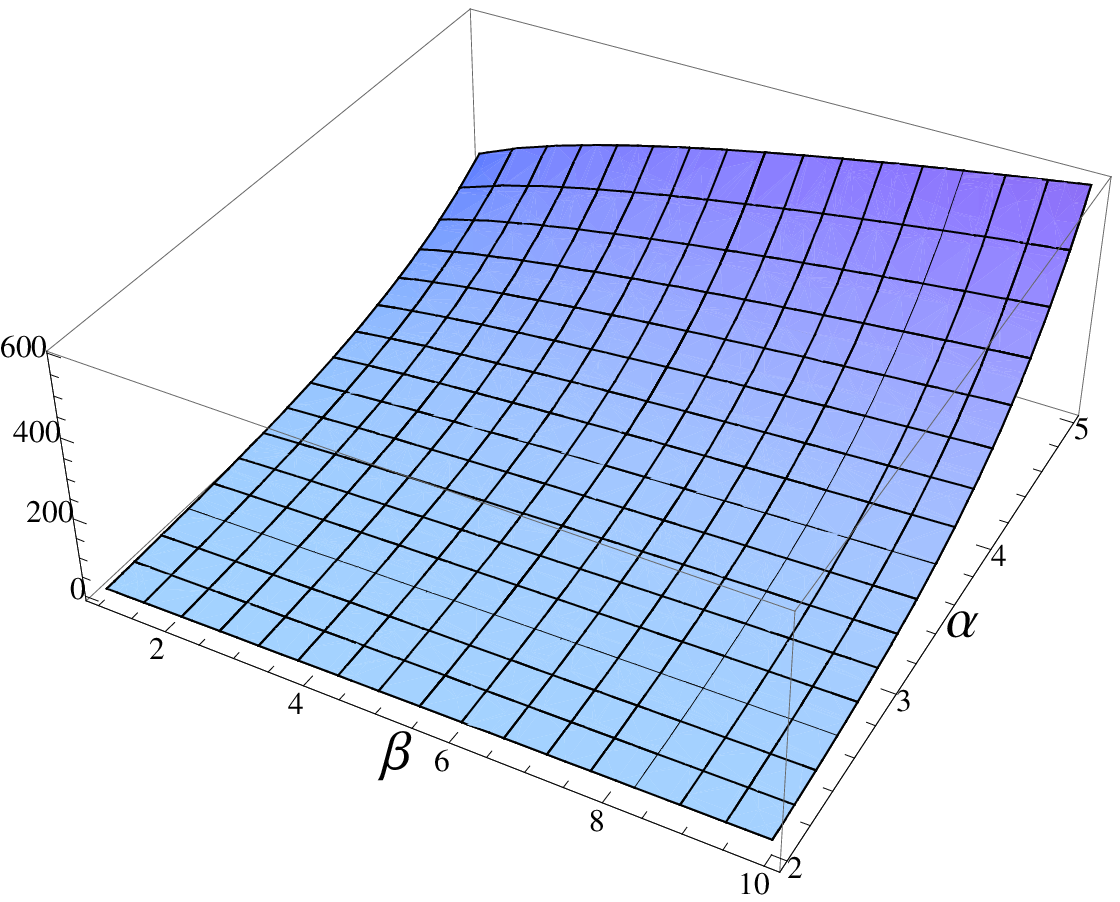}}
    } 
  \caption{Fig. (a): Numerical values of ${\sf k}(\alpha)$, which converges rapidly to finite constant $52$ when~$\alpha>2$. Fig. (b): Numerical values of $( 2 + \beta^{\frac{1}{\alpha}} )^{\alpha}$.
}\label{fig:a.k.and.b.beta} 
\end{figure}

Note that ${\sf k}(\alpha)$ converges rapidly to finite constant 52, when~$\alpha>2$. See Fig.~\ref{fig:a.k.and.b.beta}.(a) for a plot of the numerical values of ${\sf k}(\alpha)$. We remark that that \cite{FLH09safecs} considers the simpler aggregate noise-absent SIR model. Because of the absence of noise, using a tighter packing lattice \cite{FLH09safecs} yields a tighter constant ${\sf k}(\alpha)$.
\\

\begin{lemma} \label{lem:bi_fr}
If ${\sf r}'_{\sf xcl} \ge {\sf r}_{\sf xcl} + 2 {\sf r}_{\sf tx}$, then
\begin{equation}
{\mathscr U}^{\sf pw}_{\rm fr}\big[{X},{\sf r}_{\sf xcl}, {\sf r}_{\sf tx}\big] \supseteq {\mathscr B}^{\sf pw}_{\rm fr}\big[{X},{\sf r}_{\sf xcl}, {\sf r}_{\sf tx}\big] \supseteq {\mathscr U}^{\sf pw}_{\rm fr}\big[{X},{\sf r}'_{\sf xcl}, {\sf r}_{\sf tx}\big]
\end{equation}
\end{lemma}

\begin{lemma} \label{lem:bi_nf}
If $\Delta' \ge \Delta + 2$, then
\begin{equation}
{\mathscr U}^{\sf pw}_{\rm sir}\big[X,\Delta\big] \supseteq {\mathscr B}^{\sf pw}_{\rm sir}\big[X,\Delta\big] \supseteq {\mathscr U}^{\sf pw}_{\rm sir}\big[X,\Delta'\big]
\end{equation}
\end{lemma}

\begin{lemma} \label{lem:bi_pw}
If $\beta' \ge ( 2 + \beta^{\frac{1}{\alpha}} )^{\alpha}$, then
\begin{equation}
{\mathscr U}^{\sf pw}_{\rm sinr}\big[X,\beta\big] \supseteq {\mathscr B}^{\sf pw}_{\rm sinr}\big[X,\beta\big] \supseteq {\mathscr U}^{\sf pw}_{\rm sinr}\big[X,\beta'\big]
\end{equation}
\end{lemma}

\begin{lemma} \label{lem:bi_ag}
If $\beta' \ge  ( 2 + \beta^{\frac{1}{\alpha}} )^{\alpha}$, then
\begin{equation}
{\mathscr U}^{\sf ag}_{\rm sinr}\big[X,\beta\big] \supseteq {\mathscr B}^{\sf ag}_{\rm sinr}\big[X,\beta\big] \supseteq {\mathscr U}^{\sf ag}_{\rm sinr}\big[X,\beta'\big]
\end{equation}
\end{lemma}

\begin{lemma} \label{lem:cs_fr}
If ${\sf r}_{\sf tx} = \underset{i \in X}{\max} |t_i - r_i|$ and ${\sf r}_{\rm cs} \ge {\sf r}_{\sf xcl} + 2 {\sf r}_{\sf tx}$,
\begin{equation}
{\mathscr C}^{\sf pw}\big[{X},{\sf r}_{\rm xcl}\big]  \supseteq
{\mathscr B}^{\sf pw}_{\rm fr}\big[{X},{\sf r}_{\sf xcl}, {\sf r}_{\sf tx}\big] \supseteq {\mathscr C}^{\sf pw}\big[{X},{\sf r}_{\rm cs}\big]
\end{equation}
\end{lemma}

Note that Lemma~\ref{lem:bi_nf} can be proven by applying Lemma~\ref{lem:bi_pw} and letting ${\sf N_0} = 0$, $\Delta = \beta^{\frac{1}{\alpha}} - 1$ and $\Delta' = \beta'^{\frac{1}{\alpha}} - 1$. Hence, $( 2 + \beta^{\frac{1}{\alpha}} )^{\alpha}$ is a universal constant for both pairwise and aggregate interference models with/without noise. See Fig.~\ref{fig:a.k.and.b.beta}.(b) for a plot of the numerical values of $( 2 + \beta^{\frac{1}{\alpha}} )^{\alpha}$.

\subsection{Hidden-node-free Sufficient Conditions} \label{sec:subset}

\begin{figure}[htb!]
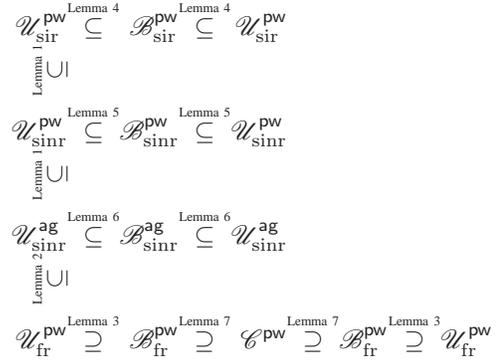
 
\begin{equation*}
\begin{array}{@{}r@{}r@{}r@{}r@{}r@{}r@{}r@{}r@{}r@{}r@{}r@{}r@{}l@{}}
{\mathscr U}^{\sf pw}_{\rm sir} & \overset{\mbox{\tiny Lemma~\ref{lem:bi_nf}}}{\subseteq} &
{\mathscr B}^{\sf pw}_{\rm sir} & \overset{\mbox{\tiny Lemma~\ref{lem:bi_nf}}}{\subseteq} &
{\mathscr U}^{\sf pw}_{\rm sir} \\
\begin{sideways}
$\overset{\mbox{\tiny Lemma~\ref{lem:pw_ag}}}{\subseteq}$
\end{sideways} \\
{\mathscr U}^{\sf pw}_{\rm sinr} & \overset{\mbox{\tiny Lemma~\ref{lem:bi_pw}}}{\subseteq} &
{\mathscr B}^{\sf pw}_{\rm sinr}  & \overset{\mbox{\tiny Lemma~\ref{lem:bi_pw}}}{\subseteq} &
{\mathscr U}^{\sf pw}_{\rm sinr} \\
\begin{sideways}
$\overset{\mbox{\tiny Lemma~\ref{lem:pw_ag}}}{\subseteq}$
\end{sideways} \\
{\mathscr U}^{\sf ag}_{\rm sinr} & \overset{\mbox{\tiny Lemma~\ref{lem:bi_ag}}}{\subseteq} &
{\mathscr B}^{\sf ag}_{\rm sinr} & \overset{\mbox{\tiny Lemma~\ref{lem:bi_ag}}}{\subseteq} &
{\mathscr U}^{\sf ag}_{\rm sinr} \\
\begin{sideways}
$\overset{\mbox{\tiny Lemma~\ref{lem:ag_fr}}}{\subseteq}$
\end{sideways} \\
{\mathscr U}^{\sf pw}_{\rm fr} & \overset{\mbox{\tiny Lemma~\ref{lem:bi_fr}}}{\supseteq} &
{\mathscr B}^{\sf pw}_{\rm fr} & \overset{\mbox{\tiny Lemma~\ref{lem:cs_fr}}}{\supseteq} &
{\mathscr C}^{\sf pw} & \overset{\mbox{\tiny Lemma~\ref{lem:cs_fr}}}{\supseteq} &
{\mathscr B}^{\sf pw}_{\rm fr} & \overset{\mbox{\tiny Lemma~\ref{lem:bi_fr}}}{\supseteq} &
{\mathscr U}^{\sf pw}_{\rm fr}
\end{array}
\end{equation*} 
 \caption{The tree diagram represents the subset-relationships for the interference and pairwise carrier-sensing decision model.} \label{fig:tree} 
\end{figure}

Lemmas~\ref{lem:pw_ag}-\ref{lem:cs_fr} establish a tree diagram Fig.~\ref{fig:tree} of subset-relationships for the interference and carrier-sensing decision models, under the respective sufficient conditions.

The tree diagram Fig.~\ref{fig:tree} provides us a way to design hidden-node-free CSMA networks. Given any bi-directional interference-safe feasible family ${\mathscr B}\big[X\big]$ from ${\sf b.0}$)-${\sf b.3}$), and pairwise carrier-sensing feasible family ${\mathscr C}^{\sf pw}\big[X, {\sf r}_{\rm cs}\big]$, we start at ${\mathscr B}\big[X\big]$ in the tree diagram, and follow the respective chains of lemmas to set the respective sufficient conditions until reaching ${\mathscr C}^{\sf pw}\big[X, {\sf r}_{\rm cs}\big]$. Then, we can obtain a hidden-node-free design. Hence, it proves the following theorem.

\begin{theorem} \label{thm:hidden}
Suppose ${\sf r}_{\sf tx} = {\max}_{i \in X} |t_i - r_i|$. For any bi-directional interference-safe feasible family ${\mathscr B}\big[X\big]$ from ${\sf b.0}$)-${\sf b.3}$) and pairwise carrier-sensing feasible family ${\mathscr C}^{\sf pw}\big[X, {\sf r}_{\rm cs}\big]$, there exists a suitable setting of ${\sf r}_{\rm cs}$ such that
\begin{equation}
(\mbox{\em Hidden-node-free Design}):
{\mathscr B}\big[X\big] \supseteq  {\mathscr C}^{\sf pw}\big[X, {\sf r}_{\rm cs}\big]
\end{equation}
\end{theorem}

We summarize the sufficient conditions for hidden-node-free CSMA network design in Table~\ref{Tab:Hidden.Node.Free.Conditions}.

We remark that although the virtual carrier sensing (RTS/CTS) in IEEE 802.11 is designed to solve the hidden node problem, using RTS/CTS in multi-hop networks does not eliminate the hidden-node problem~\cite{XGB02EffctivenessofRTSCTS}, unless the carrier sensing range is large enough and a number of other conditions are met~\cite{JL08hidden}. The conditions for hidden-node free operation under the RTS/CTS mode are much more complicated than under the basic mode, even under the pairwise interference model (see \cite{JL08hidden} for details). To keep our focus in this paper, we will not consider the RTS/CTS mode. The extension to incorporate RTS/CTS is certainly an interesting subject for future studies, particularly for the hidden-node free operation under the aggregate interference model.


\begin{table*}[hbt] 
  \centering
\begin{tabular}{c|c|c|c|c}
\hline
 & \multicolumn{4}{c}{Bi-directional feasible family}\tabularnewline
\hline
 & pairwise fixed range & pairwise SIR  & pairwise SINR & aggregate SINR\tabularnewline
 & ${\mathscr B}^{\sf pw}_{\rm fr}\big[{X},{\sf r}_{\sf xcl}, {\sf r}_{\sf tx}\big]$  & ${\mathscr B}^{\sf pw}_{\rm sir}\big[X,\Delta\big]$ & ${\mathscr B}^{\sf pw}_{\rm sinr}\big[X,\beta\big]$ & ${\mathscr B}^{\sf ag}_{\rm sinr}\big[X,\beta\big]$ \tabularnewline
\hline
\hline

\begin{tabular}{@{}c@{}}
Pairwise \\
carrier-sensing \\
feasible family \\
${\mathscr C}^{\sf pw}\big[{X},{\sf r}_{\rm cs}\big]$
\end{tabular}
&
\begin{tabular}{@{}c@{}}
${\sf r}_{\sf tx} = \underset{i \in X}{\max} |t_i - r_i|$ \\
${\sf r}_{\rm cs} \ge {\sf r}_{\sf xcl} + 2 {\sf r}_{\sf tx}$
\end{tabular}
&
\begin{tabular}{@{}c@{}}
${\sf r}_{\sf tx} = \underset{i \in X}{\max} |t_i - r_i|$ \\
${\sf r}_{\rm cs}\ge (3+\Delta){\sf r}_{\sf tx}$~\cite{JL08hidden}
\end{tabular}
&
\begin{tabular}{@{}c@{}}
${\sf r}_{\sf tx} = \underset{i \in X}{\max} |t_i - r_i|$ \\
${\sf r}_{\rm cs} \ge \Big(  \frac{1}{{\sf P}_{\sf tx} } \big( \frac{{\sf P}_{\sf tx}}{(2+\beta^{\frac{1}{\alpha}})^{\alpha}}{\sf r}_{\sf tx}^{-\alpha}$ \\
$\qquad \quad - {\sf N_0} \big) \Big)^{-\frac{1}{\alpha}} + 2{\sf r}_{\sf tx}$
~(See Lemma~\ref{lem:cs_pw_pw})
\end{tabular}
&
\begin{tabular}{@{}c@{}}
${\sf r}_{\sf tx} = \underset{i \in X}{\max} |t_i - r_i|$ \\
${\sf r}_{\rm cs} \ge \Big(  \frac{1}{{\sf P}_{\sf tx} {\sf k}(\alpha)} \big( \frac{{\sf P}_{\sf tx}}{(2+\beta^{\frac{1}{\alpha}})^{\alpha}}{\sf r}_{\sf tx}^{-\alpha}$ \\
$\qquad- {\sf N_0} \big) \Big)^{-\frac{1}{\alpha}} + 3{\sf r}_{\sf tx}$
\end{tabular}
\tabularnewline

\hline
\end{tabular} 
  \caption{Sufficient conditions for hidden-node-free CSMA network design. Results are derived in this paper unless cited otherwise.}\label{Tab:Hidden.Node.Free.Conditions} 
\end{table*}

\section{Stationary Throughput Analysis} \label{sec:random}

While Sec.~\ref{sec:models}-\ref{sec:hidden} address the distributed and ACK-based nature of CSMA, this section addresses the characteristics of random access in CSMA, and study its achievable capacity as compared to TDMA schemes.

\subsection{Deterministic Scheduling}

Consider a given routing scheme and pairwise carrier-sensing decision model ${\sf c.1}$) (implemented by IPCS and set to be hidden-node-free by Theorem~\ref{thm:hidden}). For brevity, in the following we let ${\mathscr C}\big[X\big]  \triangleq {\mathscr C}^{\sf pw}\big[X, {\sf r}_{\rm cs}\big]$. If we assume slotted time, a deterministic scheduling scheme is defined as a sequence $({\cal S}_{\sf t})_{{\sf t}={\sf 1}}^{{\sf m}}$ where each ${\cal S}_{\sf t} \in {\mathscr C}\big[X\big]$, such that the transmitters in each ${\cal S}_{\sf t}$ are allowed to transmit only at every timeslot $({\sf t} \mod {\sf m})$. A TDMA scheme is simply a deterministic scheduling scheme.  Such a TDMA scheme is only a hypothetical scheme that can serve as a ``reference'' scheme for the study of the random access based CSMA network.

Suppose the bandwidth is normalized to a unit constant. Then for each link $i \in X$, the throughput rate under scheduling scheme $({\cal S}_{\sf t})_{{\sf t}={\sf 1}}^{{\sf m}}$ is:
\begin{equation}
 {\mathfrak c}_i^{\sf det}\big[ ({\cal S}_{\sf t})_{{\sf t}={\sf 1}}^{{\sf m}} \big] \triangleq \frac{1}{\sf m} \sum_{{\sf t} = 1}^{\sf m} {\mathds 1}(i \in {\cal S}_{\sf t})
\end{equation}

Recall $\lambda_k$ is the data rate of source-sink pair $k \in N^{\sf sd}$.
With the routing scheme, one can determine the feasible region for $(\lambda_k)_{k \in N^{\sf sd}}$ by solving a multi-commodity flow problem.

\subsection{Multi-Backoff-Rate Random Access}

More generally, we consider a random access scheme (e.g., IEEE 802.11 DCF), such that $({\cal S}_{\sf t})_{{\sf t}={\sf 1}}^{{\sf \infty}}$ follows a random sequence. We consider an idealized version CSMA random access scheme as a continuous-time Markov process as in \cite{LKLW09boe,JW08csma,DDT08csma,WK05throughput}, which is sufficient to provide insights for the practical CSMA random access scheme. We assume that the count-down time and transmission time follow exponential distribution\footnote{
The main results of this paper is built upon the stationary probabilitya distribution in Eqn.~(\ref{eq:stationary.dist}). \cite{LKLW09boe} showed that for general backoff and transmission times that are not exponentially distributed, Eqn.~(\ref{eq:stationary.dist}) remains valid if the process is stationary. In particular, Eqn.~(\ref{eq:stationary.dist}) has been verified to be valid for many different backoff time distributions, including the that of Wi-Fi. Thus, strictly speaking, the exponential assumption is not needed.}.
The average count-down time can be distinct for different links. Thus, we call this {\em multi-backoff-rate random access}.
We formalize the random access scheme by a Markov chain with its states being ${\mathscr C}\big[X\big]$. There is a possible transition between states ${\cal S},{\cal S}' \in {\mathscr C}\big[X\big]$, if ${\cal S} = \{ i \} \cup {\cal S}'$ for some $i \in X$.
\begin{itemize}

\item Transition ``${\cal S} \to \{ i \} \cup {\cal S}$'' represents that the transmitter of link $i$ will start to transmit, after some random count-down time.

\item Transition ``$\{ i \} \cup {\cal S} \to {\cal S}$'' represents that the transmitter of link $i$ will finish transmission, after some random transmission time.

\end{itemize}

Suppose the current state of simultaneous transmissions is ${\cal S}$, and transmitter $i$ is counting down to transmission. 
Transmitter $i$ will freeze count-down if it detects that the channel is busy (i.e., ${\cal S} \to \{ j \} \cup {\cal S}$ for some $j \ne i$, and $\{i, j\} \cup {\cal S} \notin {\mathscr C}\big[X\big]$). It will resume count-down when the state of simultaneous transmissions becomes ${\cal S}'$ such that $\{i\} \cup {\cal S}' \in {\mathscr C}\big[X\big]$.

Let the rate of transition ${\cal S} \to \{ i \} \cup {\cal S}$ be $\nu_i$, and normalize the rate of transition $\{ i \} \cup {\cal S} \to {\cal S}$ as 1. Let ${\bf \nu} \triangleq ( \nu_i )_{i \in X}$. Then $\langle {\mathscr C}\big[X\big], {\bf \nu} \rangle$ denotes the continuous-time Markov process of idealized multi-backoff-rate CSMA random access.
\\

\begin{lemma} \label{lem:reversible}
$\langle {\mathscr C}\big[X\big], {\bf \nu} \rangle$ is a reversible Markov process, with stationary distribution for each ${\cal S} \in {\mathscr C}\big[X\big]$ as:
\begin{equation}\label{eq:stationary.dist}
{\mathbb P}_{\bf \nu}({\cal S}) = \frac{\exp\big(\sum_{i \in {\cal S}} \log \nu_i \big)}
{\sum_{{\cal S}' \in {\mathscr C}[X]}\exp\big(\sum_{j \in {\cal S}'}  \log \nu_j \big)}
\end{equation}
\end{lemma}

{\color{Black}
Lemma~\ref{lem:reversible} is well-known in the literature. We present it here for completeness. }
The long-term throughput is characterized by the stationary distribution of $\langle {\mathscr C}\big[X\big], {\bf \nu} \rangle$. Therefore, for each link $i \in X$, the throughput rate under idealized multi-backoff-rate CSMA random access is:
\begin{equation}
 {\mathfrak c}_i^{\sf rand}\big[ \langle {\mathscr C}\big[X\big], {\bf \nu} \rangle \big] \triangleq \sum_{{\cal S} \in {\mathscr C}[X]: i \in {\cal S}} {\mathbb P}_{\bf \nu}({\cal S})
\end{equation}

We can relate the throughput of a deterministic scheduling scheme with the long-term throughput of idealized multi-backoff-rate CSMA random access by the following result.
\\

\begin{lemma} \label{lem:det_rand}
Given a deterministic scheduling scheme $({\cal S}_{\sf t})_{{\sf t}={\sf 1}}^{{\sf m}}$, let the fraction of time spent in ${\cal S} \in {\mathscr C}[X]$ be ${\mathbb P}^{\sf det}({\cal S}) = \frac{1}{\sf m}\sum_{{\sf t} = 1}^{\sf m} {\mathds 1}({\cal S}_{\sf t} = {\cal S})$.
If ${\mathbb P}^{\sf det}({\cal S}) > 0$ for all ${\cal S} \in {\mathscr C}[X]$, then there exists count-down rates ${\bf \nu}$, such that for each link $i \in X$, it satisfies:
\begin{equation}
 {\mathfrak c}_i^{\sf det}\big[ ({\cal S}_{\sf t})_{{\sf t}={\sf 1}}^{{\sf m}} \big] \le {\mathfrak c}_i^{\sf rand}\big[ \langle {\mathscr C}\big[X\big], {\bf \nu} \rangle \big]
\end{equation}
\end{lemma}


Lemma~\ref{lem:det_rand} is a slightly modified version of Proposition 2 in \cite{JW08csma}, which applies to the periodic TDMA schemes as considered in this paper. In the Appendix, we give a simplified alternate proof, inspired by the set of Markov approximation arguments elaborated in \cite{CLSK10markov}. {\color{Black} A distributed algorithm is presented in \cite{JW08csma} to adapt the appropriate count-down rate ${\bf \nu}$ to satisfy Lemma~\ref{lem:det_rand}.}

The implication of Lemma~\ref{lem:det_rand} is that idealized multi-backoff-rate CSMA random access can be adapted to perform at least as well as a class of TDMA schemes under the same set of feasible states. Lemma~\ref{lem:det_rand} will be useful to explore the achievable capacity of multi-backoff-rate CSMA networks, given the achievable capacity of the corresponding TDMA scheme on the same ${\mathscr C}[X]$.

\section{Capacity of Random Network} \label{sec:net}

In this section, we apply the results from Sec.~\ref{sec:models}-\ref{sec:random} to the capacity analysis on a  uniform random network. The reason for selecting a uniform random network is to provide the simplest average-case analysis, without involving other complicated random network topologies.
We consider a Poisson point process\footnote{One can consider an alternative point process where $n$ nodes are placed on the plane by uniform distribution. But this point process converges to Poisson point process asymptotically.} of unit density on a square plane $[0, \sqrt{n}]\times[0, \sqrt{n}]$. Every node on the plane is a source or a sink that is selected uniform-randomly among all the nodes on the plane. We next define some notations:
\begin{itemize}

\item ${\cal N}^{\sf sd}_n$ denotes the {\em random} set of source-sink pairs induced by the Poisson point process.

\item ${\cal R}$ denotes a routing scheme that assigns each $k \in {\cal N}^{\sf sd}_n$ a path, such that each hop is within the maximum transmitter-receiver distance $\big({\sf P}_{\sf tx} \slash (\beta {\sf N_0})\big)^{\frac{1}{\alpha}}$.

\item ${\cal X}^{\cal R}_n$ denotes the {\em random} set of links induced by routing scheme ${\cal R}$ over ${\cal N}^{\sf sd}_n$.

\item ${\mathscr F}\big[{\cal X}^{\cal R}_n\big]$ denotes a feasible family from ${\sf a.0}$)-${\sf c.2}$) over the random set of links, ${\cal X}^{\cal R}_n$.

\item ${\mathscr S}\big({\mathscr F}\big[{\cal X}^{\cal R}_n\big]\big)$ denotes the set of all possible deterministic scheduling schemes $\big\{({\cal S}_{\sf t} \in {\mathscr F}\big[{\cal X}^{\cal R}_n\big])_{{\sf t}={\sf 1}}^{{\sf m}} \big\}$.

\item $\lambda \big( {\mathscr F}\big[{\cal X}^{\cal R}_n\big] \big)$ denotes the minimum data rate among all the source-sink pairs in ${\cal N}^{\sf sd}_n$, achieved by the optimal deterministic scheduling scheme:
\begin{equation}
\lambda \big( {\mathscr F}\big[{\cal X}^{\cal R}_n\big] \big) \triangleq
\max_{({\cal S}_{\sf t})_{{\sf t}={\sf 1}}^{{\sf m}} \in {\mathscr S}({\mathscr F}[{\cal X}^{\cal R}_n])}
\Big( \min_{k \in {\cal N}^{\sf sd}_n} \lambda_k \Big)
\end{equation}

\end{itemize}

We now define the capacity over random networks. Since $\lambda \big( {\mathscr F}\big[{\cal X}^{\cal R}_n\big] \big)$ is a random variable, we say that the capacity over ${\cal N}^{\sf sd}_n$ has an order as $\Theta(f(n))$ with high probability (w.h.p.), if there exists finite constants $c' > c > 0$ such that 
\[
\begin{array}{@{}r@{}l@{}r}
\lim_{n \to \infty} & {\mathbb P}\big\{ \lambda \big( {\mathscr F}\big[{\cal X}^{\cal R}_n\big] \big) = c \cdot f(n) \mbox{\ is feasible}\big\} & = 1 \\
\liminf_{n \to \infty} & {\mathbb P}\big\{ \lambda \big( {\mathscr F}\big[{\cal X}^{\cal R}_n\big] \big) = c' \cdot f(n) \mbox{\ is feasible}\big\} & < 1
\end{array}
\]
This is the conventional definition of random wireless network capacity \cite{GK00capacity,XK06scale,FDTT00percolation,LLL08multicast}.

\subsection{Upper Bound for Single Carrier Sensing}
We first show that carrier sensing based on ${\sf c.1}$)-${\sf c.2}$) cannot achieve the optimal capacity ${\Omega}\big(\frac{1}{\sqrt{n}}\big)$.

\begin{theorem} \label{thm:bd_single}
Consider a carrier-sensing feasible family ${\mathscr C}\big[{\cal X}^{\cal R}_n\big]$ from ${\sf c.1}$)-${\sf c.2}$), for any routing scheme ${\cal R}$ that connects all the source-sink pairs in ${\cal N}^{\sf sd}_n$,
\begin{equation}
\lambda \big( {\mathscr C}\big[{\cal X}^{\cal R}_n\big] \big) = {\rm O}\Big(\frac{1}{\sqrt{n \log n}}\Big) \mbox{\quad (w.h.p.)}
\end{equation}
\end{theorem}

\begin{proof}
By Lemmas~\ref{lem:bi_fr},\ref{lem:cs_fr}, there exists a suitable ${\sf r}_{\sf xcl}$, such that ${\mathscr C}\big[{\cal X}^{\cal R}_n\big]$ can be configured as a subset of ${\mathscr U}^{\sf pw}_{\rm fr}\big[{\cal X}^{\cal R}_n,{\sf r}_{\sf xcl}, {\sf r}_{\sf tx}\big]$. It has been shown in \cite{GK00capacity} that
\begin{equation}
\lambda \big( {\mathscr U}^{\sf pw}_{\rm fr}\big[{\cal X}^{\cal R}_n,{\sf r}_{\sf xcl}, {\sf r}_{\sf tx}\big] \big) = {\rm O}\Big(\frac{1}{\sqrt{n \log n}}\Big) \mbox{\quad (w.h.p.)}
\end{equation}
for any routing scheme ${\cal R}$ that connects all the source-sink pairs in ${\cal N}^{\sf sd}_n$.
Hence, it completes the proof.
\end{proof}

Nonetheless, \cite{FDTT00percolation,XK06scale} show that for any interference-safe feasible family from ${\sf a.1}$)-${\sf a.3}$), there exists a TDMA scheme to achieve throughput as ${\Omega}\big(\frac{1}{\sqrt{n}}\big)$ (w.h.p.). We are thus motivated to adopt such a TDMA-based approach to CSMA networks.

\subsection{Backbone-Peripheral Routing} \label{sec:routing}

For the completeness of presentation, we briefly revisit the efficient routing scheme in \cite{FDTT00percolation} (we call {\em backbone-peripheral routing}). Partition the nodes into two classes: {\em backbone nodes} and {\em peripheral nodes}. The backbone nodes themselves are connected using only short-range links, whereas every peripheral node can reach a backbone node in one-hop transmission. The basic idea is to use short-range backbone-backbone links whenever possible. Since short-range links generate minimal spatial interference, this increases the number of simultaneous active links, and hence the throughput.

To implement backbone-peripheral routing, we first partition the square plane $[0, \sqrt{n}]\times[0, \sqrt{n}]$ into square cells with sidelength ${\sf s}_n$. Consider the cells as vertices, a path can be formed by connecting adjacent non-empty cells.
\\

\begin{lemma} \label{lem:percolate} (See \cite{FDTT00percolation})
There exist constants $c_1, c_2, c_3$ independent of $n$, such that when we set ${\sf s}_n = c_1$, then in every horizontal slab of $(\sqrt{n}/c_1 \times c_2 \log n/c_1)$ cells, there exist at least $c_3 \log n$ disjoint paths between the vertical opposite sides of the plane (w.h.p.).
\end{lemma}
\

We build a backbone (called ``highway system'' in \cite{FDTT00percolation}) for routing on a uniform random network as follows. Select a representative node in each non-empty cell. By Lemma~\ref{lem:percolate}, there is a connected sub-network that spans the plane (w.h.p.), formed by connecting the representative nodes in the adjacent cells. These connected representative nodes are the {\em backbone nodes}, while the rest are the {\em peripheral nodes}. Note that the distance between two adjacent backbone nodes is at most $\sqrt{5} c_1$, while the distance between a peripheral node to a nearby backbone node is at most $c_2 \log n$ (w.h.p.).

\begin{figure}[htb!] 
    \centering
      \includegraphics[scale=0.5]{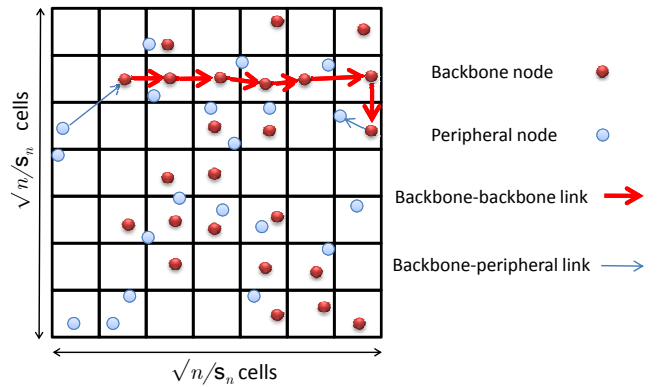} 
  \caption{Backbone nodes are a subset of connected nodes by short-range links, whereas peripheral nodes relay all the packets to backbone nodes.} \label{fig:bk-ph} 
\end{figure}

Backbone-peripheral routing scheme operates as follows. The source first uses a one-hop transmission to a backbone node, if it is a peripheral node. We control the packet load from the peripheral nodes such that each backbone node is accessed by at most by some constant number of peripheral nodes. Next, the receiving backbone node relays the packet following multi-hop Manhattan-routing along the adjacent backbone nodes to the respective backbone node that can transmit the packets to the sink in a single last hop. See Fig.~\ref{fig:bk-ph} for an illustration of backbone-peripheral routing.

We define a scheduling scheme under backbone-peripheral routing consisting of two stages: $({\cal S}_{\sf t}^{\sf P})_{{\sf t}={\sf 1}}^{c_4 \log^2 n}$ and $({\cal S}_{\sf t}^{\sf B})_{{\sf t}={\sf 1}}^{c_5}$, for some constants $c_4, c_5$.
\begin{enumerate}

\item ({\em Backbone-peripheral Transmissions}):
If $i \in {\cal S}_{\sf t}^{\sf P}$, then either $t_i$ or $r_i$ is a peripheral node. Using a spatial assignment scheme, we divide the plane into larger cells, each of which having an area of $\Theta(\log^2 n)$ (because the backbone-peripheral distance is ${\rm O}(\log n)$). It is shown in \cite{FDTT00percolation} that we can always pick a non-interfering link in each cell to transmit in every timeslot $({\sf t} \mod c_4 \log^2 n)$ in the first stage, for some constant $c_4$. The throughput rate for each backbone-peripheral link can be shown to be $\Theta(\frac{1}{\log^2 n}) \gg \Theta(\frac{1}{\sqrt{n}})$.

\item  ({\em Backbone-backbone Transmissions}):
If $i \in {\cal S}_{\sf t}^{\sf B}$, then both $t_i$ and $r_i$ are backbone nodes. Since the backbone-backbone distance is ${\rm O}(1)$, we use a similar spatial assignment scheme but considering a cell with an area $c_5$, for some constant $c_5$. Since each backbone node is accessed by at most by some constant number of peripheral nodes, there are at most ${\rm O}(\sqrt{n})$ peripheral nodes that relays packets to each backbone node. Thus, the throughput rate at each backbone-backbone link divided by the number of peripheral nodes that relay packets to it is $\Theta(\frac{1}{\sqrt{n}})$.

\end{enumerate}
Overall, backbone-backbone links are the bottleneck, not backbone-peripheral links. Hence, $\lambda_k = {\Omega}(\frac{1}{\sqrt{n}})$ is achievable w.h.p. on a uniform random network based on backbone-peripheral routing and the above two-stage scheduling scheme.

{\color{Black}
Note that since the maximum backbone-peripheral distance may scale as ${\Theta}(\log n)$, it is necessary to decrease threshold $\beta$ or increase power ${\sf P_{tx}}$ as $n$ increases for these links in the SINR models. If we opt to keep a fixed ${\sf P_{tx}}$ and decrease $\beta$ , the data rate will decrease as $n$ increases. However, the data rate does not decrease as fast as the target per-flow throughput, which is ${\rm O}(\frac{1}{\sqrt{n}})$. Thus, the bottleneck will remain to be at backbone-backbone links. }

\section{Dual Carrier-Sensing} \label{sec:multi}

To adopt the TDMA scheme of backbone-peripheral routing in Sec.~\ref{sec:routing} for CSMA networks, in this section we employ {\em dual carrier-sensing} where multiple carrier-sensing ranges are allowed. Namely, smaller carrier-sensing ranges can be used among the short-range links. This effectively enables more simultaneous links and improves the throughput.

However, it is not straightforward to implement dual carrier-sensing in conventional CSMA protocols (e.g., IEEE 802.11), because the transmitters may not be aware if the other active links are short-range or long-range. To address the above implementation issue of dual carrier-sensing, we are motivated to adopt a system with two frequency channels, in which the communications on the backbone-backbone links are carried out on one frequency channel, while the communications on the peripheral links are carried out on the other channel.

In the following, we provide a detailed study on the implementation of dual carrier-sensing on two frequency channels. First, Sec. \ref{subsec:fulldup} considers a system that is full-duplex across the two frequency channels. Then, Sec. \ref{subsec:halfdup} considers a system that is half-duplex across the two frequency channels that is simpler to implement, but whose conditions for hidden-node free operation are more subtle.

\subsection{Full-duplexity across Two Frequency Channels} \label{subsec:fulldup}

Thus far, we have assumed that the communication on a channel is half-duplex in that when a node transmits, it cannot receive. This is typically the case if one strives for simple transceiver designs. We will continue to assume that a node cannot transmit and receive on the same channel simultaneously. However, we assume {\em full-duplexity} across different frequency channels in that simultaneous transmission and reception on different channels are allowed.  Specifically, when a node transmits on frequency 1, it could receive on frequency 2; and when a node transmits on frequency 2, it could receive on frequency 1.
\\

{\em Carrier-sening Mechanism:}
With such set-up, the peripheral nodes will transmit and receive on one of the frequency channels, referred to as the {\em peripheral channel}. The backbone nodes will transmit and receive among themselves on the backbone subnet using the other frequency channel, referred to as the {\em backbone channel}. When transmitting to or receiving from the peripheral nodes, however, the backbones nodes will use the peripheral channel. Thus, a backbone node can conceptually be thought of as consisting of two virtual nodes: a virtual peripheral node for communicating with peripheral nodes associated with it; and a virtual backbone node for relaying packets over the backbone network. This design decouples the operation of the peripheral access subnet from that of the backbone highway.

Formally, we partition ${X}$ into two disjoint classes: ${X}^{\sf B}$ for backbone-backbone links, and ${X}^{\sf P}$ for backbone-peripheral links. Assume ${\sf r}_{\rm cs}^{\sf B} < {\sf r}_{\rm cs}^{\sf P}$. The feasible family that captures the above carrier-sensing mechanism is defined as:
{\color{Black}
\begin{enumerate}

\item[${\sf d.1}$)]
{\em Full-duplex pairwise dual carrier-sensing feasible family}:
${\cal S} \in {\mathscr C}^{\sf pw}_{\rm ful}\big[ ({X}^{\sf B}, {\sf r}_{\rm cs}^{\sf B}), ({X}^{\sf P}, {\sf r}_{\rm cs}^{\sf P})\big]$, if and only if for all $i, j \in {\cal S}$,
\begin{equation}
|t_j - t_i| \ge {\sf r}_{\rm cs}^{\sf c}
\end{equation}
such that $i, j \in {X}^{\sf c}$ and ${\sf c} \in \{ {\sf B}, {\sf P}\}$.

\end{enumerate}
That is, a peripheral node will carrier-sense the peripheral channel only. A backbone node will carrier-sense the peripheral channel if it wishes to transmit to a peripheral node, and will carrier-sense the backbone channel if it wishes to transmit to a backbone node.
}
\\

{\em Throughput:}
We now show that carrier-sensing model ${\sf d.1}$) can achieve throughput as ${\Omega}\big(\frac{1}{\sqrt{n}}\big)$ on two independent frequency channels.
\\

\begin{theorem} \label{thm:bd_dual}
Consider full-duplex pairwise dual carrier-sensing model ${\sf d.1}$) on a uniform random network based on backbone-peripheral routing. Let ${\cal X}^{\sf B}_n$ and ${\cal X}^{\sf P}_n$ be the random set of induced backbone-backbone links and backbone-peripheral links, respectively. Using multi-backoff-rate random access scheme, there exists a suitable setting of $({\sf r}_{\rm cs}^{\sf B}, {\sf r}_{\rm cs}^{\sf P})$, such that
\begin{equation}
\lambda \big( {\mathscr C}^{\sf pw}_{\rm ful}\big[ ({X}^{\sf B}, {\sf r}_{\rm cs}^{\sf B}), ({X}^{\sf P}, {\sf r}_{\rm cs}^{\sf P})\big] \big) = {\Omega}\Big(\frac{1}{\sqrt{n}}\Big) \mbox{\quad (w.h.p.)}
\end{equation}
\end{theorem}

\begin{proof}
Recall that $({\cal S}_{\sf t}^{\sf P})_{{\sf t}={\sf 1}}^{c_4 \log^2 n}$ and $({\cal S}_{\sf t}^{\sf B})_{{\sf t}={\sf 1}}^{c_5}$ are the two stage TDMA schemes in backbone-peripheral routing.
Note that each of ${\cal S}_{\sf t}^{\sf P}$ and ${\cal S}_{\sf t}^{\sf B}$ is a feasible state in some uni-directional pairwise fixed-range feasible families ${\sf a.0}$), where the transmitter-receiver distance is ${\rm O}(\log n)$ and ${\rm O}(1)$ respectively. By Lemma~\ref{lem:bi_fr}, we can obtain bi-directional feasible versions of $({\cal S}_{\sf t}^{\sf P})_{{\sf t}={\sf 1}}^{c_4 \log^2 n}$ and $({\cal S}_{\sf t}^{\sf B})_{{\sf t}={\sf 1}}^{c_5}$, denoted as $({\cal S'}_{\sf t}^{\sf P})_{{\sf t}={\sf 1}}^{c_4 \log^2 n}$ and $({\cal S'}_{\sf t}^{\sf B})_{{\sf t}={\sf 1}}^{c_5}$, which can be applied to backbone-peripheral routing scheme without altering the order results on capacity.

{\color{Black}
Since the two frequency channels are independent, by Theorem~\ref{thm:hidden}, we can obtain suitable settings of ${\sf r}_{\rm cs}^{\sf B}$ and ${\sf r}_{\rm cs}^{\sf P}$, such that ${\mathscr C}^{\sf pw}\big[ {X}^{\sf B}, {\sf r}_{\rm cs}^{\sf B}]$ and ${\mathscr C}^{\sf pw}\big[ {X}^{\sf P}, {\sf r}_{\rm cs}^{\sf P}]$ are hidden node free in their respective channels with respect to any interference model ${\sf b.0})$-${\sf b.2})$, and
\begin{equation}
{\cal S'}_{\sf t}^{\sf P} \subseteq {\mathscr C}^{\sf pw}\big[ {X}^{\sf B}, {\sf r}_{\rm cs}^{\sf B}]
\mbox{\ for all\ } {\sf t}={\sf 1} ... {c_4 \log^2 n}
\end{equation}
\begin{equation}
{\cal S'}_{\sf t}^{\sf B} \subseteq {\mathscr C}^{\sf pw}\big[ {X}^{\sf P}, {\sf r}_{\rm cs}^{\sf P}]
\mbox{\ for all\ } {\sf t}={\sf 1} ... {c_5}
\end{equation}
}

Next, we employ Lemma~\ref{lem:det_rand} to establish a lower bound of the throughput of random access on each of ${\mathscr C}^{\sf pw}\big[ {X}^{\sf B}, {\sf r}_{\rm cs}^{\sf B}]$ and ${\mathscr C}^{\sf pw}\big[ {X}^{\sf P}, {\sf r}_{\rm cs}^{\sf P}]$, by the throughput of a corresponding deterministic scheduling scheme as follows:
\begin{itemize}

\item
For each ${\cal S} \in \{ {\cal S'}_{\sf t}^{\sf P} \}_{{\sf t}={\sf 1}}^{{c_4 \log^2 n}}$,
we set ${\mathbb P}^{\sf det}({\cal S}) = \Theta({\frac{1}{\log^2 n}})$

\item
For each ${\cal S} \in \{ {\cal S'}_{\sf t}^{\sf B} \}_{{\sf t}={\sf 1}}^{{c_5}}$,
we set ${\mathbb P}^{\sf det}({\cal S}) = \Theta(1)$

\end{itemize}
Therefore, this satisfies the sufficient condition in Lemma~\ref{lem:det_rand} that ${\mathbb P}^{\sf det}({\cal S}) > 0$. It is easy to see that ${\mathscr C}^{\sf pw}_{\rm ful}\big[ ({X}^{\sf B}, {\sf r}_{\rm cs}^{\sf B}), ({X}^{\sf P}, {\sf r}_{\rm cs}^{\sf P})\big]$ is just a product of ${\mathscr C}^{\sf pw}\big[ {X}^{\sf B}, {\sf r}_{\rm cs}^{\sf B}]$ and ${\mathscr C}^{\sf pw}\big[ {X}^{\sf B}, {\sf r}_{\rm cs}^{\sf B}]$. Since such a deterministic scheduling scheme can achieve throughput as ${\Omega}\big(\frac{1}{\sqrt{n}}\big)$ on a uniform random network w.h.p., it completes the proof by Lemma~\ref{lem:det_rand}.
\end{proof}

\subsection{Half-duplexity across Two Frequency Channels} \label{subsec:halfdup}

We now consider a system that is {\em half-duplex} across the two frequency channels to ease implementation
further. A node can still receive on different channels simultaneously (for the purpose of carrier-sensing
both channels simultaneously rather than receiving data targeted for it). However, we place a restriction
on  simultaneous transmission and reception on the same channel or different channels, as elaborated below.
\\

{\em Half-duplexity Constraints:} We introduce the following constraints to formulate half-duplexity:
\begin{itemize}

\item[({\em i})] a node cannot transmit on channel $i$ and receive on channel $j$ at the same time,
whether $i=j$ or $i \neq j$.

\item[({\em ii})] a node can only transmit on at most one frequency channel at any time.

\end{itemize}

Constraint ({\em i}) is mainly to simplify implementation. When a node transmits, its own transmitted signal power may overwhelm the received signal.
Although in principle, the use of a frequency filter may be able to isolate the signals somewhat, the transmit power may be very large compared with the receive power (i.e., extreme near-far problem), such that leakage or crosstalk from the power at the transmit band may not be negligible compared with the receive power. Reference \cite{ASH08nonequ} contains a discussion on the need for the assumption of half-duplexity when the transmit and receive frequency channels are the same, but the underlying rationale and principles are the same when the cross-frequency leakage is not negligible.

Constraint ({\em ii}) is mainly due to the fact that in ACK-based CSMA schemes (e.g., IEEE 802.11), there is an ACK packet in the reverse direction after the transmission of a DATA packet. If the nodes transmit on two frequency channels and the DATA packets are of different lengths, one of the DATA frames may finish first and the station may end up transmitting DATA and receiving ACK packets at the same time, thus violating constraint ({\em i}).
\\

\begin{figure}[htb!] 
    \centering
      \includegraphics[scale=0.5]{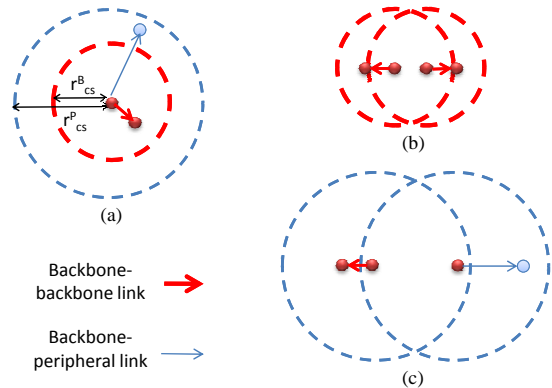} 
  \caption{There are two carrier sensing ranges as in Fig. (a). In Fig. (b) short-range backbone-backbone links will use a shorter carrier-sensing range among themselves, while in Fig. (c) longer carrier-sensing range will used when there is any active backbone-peripheral link.} \label{fig:mcs} 
\end{figure}

{\em Carrier-sening Mechanism:}
We now describe the carrier-sening mechanism under constraints ({\em i}) and ({\em ii}) as follows. The mechanism is illustrated in Fig.~\ref{fig:mcs}. The basic idea is that we allow a shorter carrier-sensing range to be used among backbone-backbone links, whereas a longer carrier-sensing range to be used in both channels when there is an active backbone-peripheral link in the neighborhood.

First, we consider the case of a backbone-peripheral link, where its carrier-sensing range is ${\sf r}_{\rm cs}^{\sf P}$. In this case, either a peripheral node desires to transmit to a backbone node, or a backbone node desires to transmit to a peripheral node. The transmission cannot be allowed if there is any simultaneous transmitters within the carrier-sensing range ${\sf r}_{\rm cs}^{\sf P}$ in either peripheral channel or backbone channel. Since ${\sf r}_{\rm cs}^{\sf P} \ge {\sf r}_{\rm cs}^{\sf B}$, this implies precluding transmission in backbone channel under carrier-sensing range~${\sf r}_{\rm cs}^{\sf B}$.

The reason for this requirement is because of the following consideration. Suppose that a peripheral node wants to transmit to its access backbone node. It is possible that the backbone node is in the midst of a communication with another backbone node. To make sure that the peripheral node does not initiate a transmission to the backbone node in that situation, the peripheral node also has to perform carrier-sensing on the backbone channel. In practice, further implementation optimization is possible (skipped here due to limited space).


Next, we consider the case of a backbone-backbone link, where its carrier-sensing range is ${\sf r}_{\rm cs}^{\sf B}$. In this case, a backbone node wants to transmit to another backbone node.  The transmission is not allowed if 1) there is any simultaneous transmitters within the carrier-sensing range ${\sf r}_{\rm cs}^{\sf B}$ in the backbone channel, or 2) there is any simultaneous transmitters within the carrier-sensing range ${\sf r}_{\rm cs}^{\sf P}$ in the peripheral channel. The former condition is obvious. The latter condition is due to the fact that the target receiver backbone node may be in the midst of a communication with a peripheral node. Again, further optimization is possible with the latter case. Here, we simply set the carrier-sensing range in the later to be ${\sf r}_{\rm cs}^{\sf P}$, since the order results we want to establish are not compromised.

The feasible family that captures the above carrier-sensing mechanism is defined as:
{\color{Black}
\begin{enumerate}

\item[${\sf d.2}$)]
{\em Half-duplex pairwise dual carrier-sensing feasible family}:
${\cal S} \in$
 ${\mathscr C}^{\sf pw}_{\rm haf}\big[ ({X}^{\sf B}, {\sf r}_{\rm cs}^{\sf B}), ({X}^{\sf P}, {\sf r}_{\rm cs}^{\sf P})\big]$, if and only if for all $i,j \in {\cal S}$,
\begin{equation}
|t_j - t_i| \ge \max\{ {\sf r}_{\rm cs}^{\sf c}, {\sf r}_{\rm cs}^{\sf c'}\},
\end{equation}
where $i \in {X}^{\sf c}, j \in {X}^{\sf c'}$ and ${\sf c}, {\sf c'} \in \{ {\sf B}, {\sf P}\}$.

\end{enumerate}
That is, there is a dynamic switching process of carrier-sensing ranges, depending on the presence of the classes of active links.
}
\\

{\em Throughput:}
Since we are considering half-duplexity across two frequency channels, the proof of throughput is different than Theorem~\ref{thm:bd_dual}. To show that carrier-sensing model ${\sf d.2}$) in the presence of half-duplexity can achieve the throughput as ${\Omega}\big(\frac{1}{\sqrt{n}}\big)$, we first need to determine ${\sf r}_{\rm cs}^{\sf P}$ and ${\sf r}_{\rm cs}^{\sf B}$. We have to formally show carrier-sensing decision model ${\sf d.2}$) can be implemented practically, by considering dual channel interference models that explicitly incorporate the constraint of half-duplexity across two frequency channels. Thus, we define the aggregate interference model in such case as follows:
\begin{enumerate}

\item[${\sf e.1}$)]
\mbox{\em Half-duplex bi-directional dual channel aggregate SINR}

{\em feasible family}:
${\cal S} \in$ ${\mathscr B}^{\sf ag}_{\rm haf}\big[({X}^{\sf B}, \beta), ({X}^{\sf P}, \beta) \big]$, if and only if

\begin{enumerate}

\item[${\sf 1}$)]
 ${\cal S} = \underset{{\sf c} \in \{ {\sf B}, {\sf P}\}}{\bigcup} {\cal S}^{\sf c}$, where each ${\cal S}^{\sf c} \in {\mathscr B}^{\sf ag}_{\rm sinr}\big[{X}^{\sf c},\beta\big]$,

\item[${\sf 2}$)] (half-duplexity constraint)
for any pair $i, j \in {\cal S}$,
$\{t_i, r_i \} \cap \{t_j, r_j \} = \varnothing$.

\end{enumerate}

\end{enumerate}
Similarly, one can define the respective dual channel interference models for ${\sf a.0})$-${\sf a.3})$,${\sf b.0})$-${\sf b.2})$.
\\

{\color{Black}
\begin{theorem} \label{thm:mbi_mcs}
There exists a suitable setting of $({\sf r}_{\rm cs}^{\sf B}, {\sf r}_{\rm cs}^{\sf P})$, depending on $\beta$ and the maximum transmission distance in ${X}^{\sf c}$, such that
\[
{\mathscr B}^{\sf ag}_{\rm haf}\big[({X}^{\sf B}, \beta), ({X}^{\sf P}, \beta) \big] \supseteq  {\mathscr C}^{\sf pw}_{\rm haf}\big[({X}^{\sf B}, {\sf r}_{\rm cs}^{\sf B}), ({X}^{\sf P}, {\sf r}_{\rm cs}^{\sf P})\big]
\]
\end{theorem}
}
\

Theorem~\ref{thm:mbi_mcs} establishes a hidden-node-free design for the dual carrier-sensing decision model. The proof of Theorem~\ref{thm:mbi_mcs} is to apply the single-channel hidden-node-free design (Theorem~\ref{thm:hidden}) on two independent frequency channels, and then show the half duplexity constraint in ${\sf e.1}$) will not affect the setting of hidden-node-free design in ${\sf d.2}$).

Similar to Theorem~\ref{thm:bd_dual}, by Theorem~\ref{thm:mbi_mcs} we can immediately show that ${\sf d.2}$) can also achieve throughput as ${\Omega}\big(\frac{1}{\sqrt{n}}\big)$.
\\

\begin{theorem} \label{thm:bd_dmcs}
Consider half-duplex pairwise dual carrier-sensing model ${\sf d.2}$) on a uniform random network based on backbone-peripheral routing. Using multi-backoff-rate random access, there exists a suitable setting of $({\sf r}_{\rm cs}^{\sf B}, {\sf r}_{\rm cs}^{\sf P})$, such that
\begin{equation}
\lambda \big( {\mathscr C}^{\sf pw}_{\rm haf}\big[ ({X}^{\sf B}, {\sf r}_{\rm cs}^{\sf B}), ({X}^{\sf P}, {\sf r}_{\rm cs}^{\sf P})\big] \big) = {\Omega}\Big(\frac{1}{\sqrt{n}}\Big) \mbox{\quad (w.h.p.)} 
\end{equation}
\end{theorem}
\

We remark that our CSMA capacity scaling-law results also hold for dense networks (where all $n$ nodes are packed in a fixed area $[0,1]\times[0,1]$), because the construction of backbone-peripheral routing also applies to dense networks \cite{FDTT00percolation}.

\section{Conclusion}\label{sec:Conclus}

This paper contains a number of new results and ideas that lend insights and solutions to maximize
the achievable capacity in CSMA wireless networks.
We formulate a comprehensive set of CSMA models, considering various distributed decision controls and common interference settings from the literature. We establish the relationship between our CSMA models with the existing interference models from the literature. This can characterize both the upper and achievable bounds on the capacity of CSMA networks to be $\Theta(\frac{1}{\sqrt{n}})$.

We show that, based on an efficient backbone-peripheral routing scheme and a careful design of dual
carrier-sensing and dual channel scheme, hidden-node-free CSMA networks can achieve throughput as $\Omega(\frac{1}{\sqrt{n}})$, as optimal as
TDMA schemes can on a uniform random network. Along the journey,
we also show that normal, single, and homogeneous carrier sensing operation is insufficient to achieve the capacity as optimal as TDMA schemes can on a uniform random network.


\section*{Acknowledgment}
\addcontentsline{toc}{section}{Acknowledgment}

The authors would like to thank Bin Tang, Libin Jiang, Yang Yang and Kit Wong for helpful comments and suggestions.



\section{Appendix}

\setcounter{lemma}{10}

\begin{lemma} \label{lem:ineq1}
For $\gamma > 1$, it is straightforward that
\begin{equation}
\displaystyle \frac{A}{{\sf N_0} + \gamma B} < \beta
\ \Rightarrow \ \displaystyle \frac{A}{{\sf N_0} +  B} < \gamma \beta
\end{equation}
\end{lemma}

\begin{lemma} \label{lem:onion}
Let ${\sf r}_{\sf tx} = {\max}_{i \in X} |t_i - r_i|$. If there exists ${\sf r}_{\sf xcl} > {\sf r}_{\sf tx}$ such that $|t_j - r_i| \ge {\sf r}_{\sf xcl}$ for all $i, j \in {\cal S}$, then
\begin{equation}
\sum_{j \in {\cal S} \backslash \{ i \}} |t_j - r_i|^{-\alpha} \le {\sf k}(\alpha) ({\sf r}_{\sf xcl}-{\sf r}_{\sf tx})^{-\alpha}
\end{equation}
where ${\sf k}(\alpha) \triangleq \sum_{k=1}^{\infty} 4 \lceil \pi (2k + 2) \rceil k^{-\alpha}$.
\end{lemma}
\begin{proof}
The proof is adopted from \cite{LLL08multicast} (Lemma 3). 

For all $i, j \in {\cal S}$, we have
\begin{equation} \label{eq:onion_d}
|t_i - t_j| \ge {\sf r}_{\sf xcl}-{\sf r}_{\sf tx}
\end{equation}
For any non-negative integer $k$, let 
\begin{equation}
T_k(r_i) = \Big\{ j \in X \backslash \{i\} \mid k ({\sf r}_{\sf xcl}-{\sf r}_{\sf tx}) \le |t_j - r_i| < (k+1)({\sf r}_{\sf xcl}-{\sf r}_{\sf tx}) \Big\}
\end{equation}

By Eq.~(\ref{eq:onion_d}), we obtain\footnote{Note that \cite{LLL08multicast} (Lemma 3) contains a slight mistake as $|T_k(r_i)| \le 2 \lceil \pi (2k + 2) \rceil$.}
\begin{equation}
|T_k(r_i)| \le 4 \lceil \pi (2k + 2) \rceil
\end{equation}

Hence, it follows that
\begin{equation}
\begin{array}{r@{\ }l}
\displaystyle \sum_{j \in {\cal S} \backslash \{ i \}} |t_j - r_i|^{-\alpha} \le 
& \displaystyle \sum_{k=1}^{\infty} |T_k(r_i)| \big(k \cdot ({\sf r}_{\sf xcl}-{\sf r}_{\sf tx}) \big)^{-\alpha} \\
\le & \displaystyle \sum_{k=1}^{\infty} 4 \lceil \pi (2k + 2) \rceil\big(k \cdot ({\sf r}_{\sf xcl}-{\sf r}_{\sf tx}) \big)^{-\alpha}
\end{array}
\end{equation}
\end{proof}

\begin{corollary} \label{cor:onion}
By Lemma~\ref{lem:onion}, if there exists ${\sf r}_{\sf cs}$ such that $|t_j - t_i| \ge {\sf r}_{\sf cs}$ for all $i, j \in {\cal S}$, then
\begin{equation}
\begin{array}{c}
\sum_{j \in {\cal S} \backslash \{ i \}} |t_j - t_i|^{-\alpha} \le {\sf k}(\alpha) ({\sf r}_{\sf cs})^{-\alpha}
\end{array}
\end{equation}
\end{corollary}

\setcounter{lemma}{0}

\begin{lemma}
If $\Delta \le \beta^{\frac{1}{\alpha}}-1$, then
\begin{equation}
{\mathscr U}^{\sf pw}_{\rm sir}[X,\Delta] \supseteq {\mathscr U}^{\sf pw}_{\rm sinr}[X,\beta]
\supseteq {\mathscr U}^{\sf ag}_{\rm sinr}[X,\beta]
\end{equation}
\end{lemma}
\begin{proof}
${\mathscr U}^{\sf pw}_{\rm sinr}[X,\beta] \supseteq {\mathscr U}^{\sf ag}_{\rm sinr}[X,\beta]$ is trivial.

${\mathscr U}^{\sf pw}_{\rm sir}[X,\Delta] \supseteq {\mathscr U}^{\sf pw}_{\rm sinr}[X,\beta]$ follows from:
\begin{equation}
\begin{array}{r@{\ }l}
 \frac{{\sf P}_{\sf tx} |t_i - r_i|^{-\alpha}}{{\sf N_0} +
 {\sf P}_{\sf tx}  |t_j - r_i|^{-\alpha}} \ge \beta
 & \Rightarrow\
 \frac{{\sf P}_{\sf tx} |t_i - r_i|^{-\alpha}}{
 {\sf P}_{\sf tx}  |t_j - r_i|^{-\alpha}} \ge \beta  \\
  & \Rightarrow\
 |t_j - r_i| \ge \beta^{\frac{1}{\alpha}} |t_i - r_i|
\end{array}
\end{equation}
\end{proof}

\begin{lemma}
Let ${\sf r}_{\sf tx} = {\max}_{i \in X} |t_i - r_i|$. If
\begin{equation}
{\sf r}_{\sf xcl} \ge
\Big(  \frac{1}{{\sf P}_{\sf tx} {\sf k}(\alpha)} \big( \frac{{\sf P}_{\sf tx}}{\beta}{\sf r}_{\sf tx}^{-\alpha} - {\sf N_0} \big) \Big)^{-\frac{1}{\alpha}} + {\sf r}_{\sf tx}
\end{equation}
where ${\sf k}(\alpha) \triangleq \sum_{k=1}^{\infty} 4 \lceil \pi (2k + 2) \rceil k^{-\alpha}$,
then
\begin{equation}
{\mathscr U}^{\sf ag}_{\rm sinr}[X,\beta] \supseteq  {\mathscr U}^{\sf pw}_{\rm fr}\big[{X},{\sf r}_{\sf xcl}, {\sf r}_{\sf tx}\big]
\end{equation}
\end{lemma}
\begin{proof}
Suppose ${\cal S} \in {\mathscr U}^{\sf pw}_{\rm fr}\big[{X},{\sf r}_{\sf xcl}, {\sf r}_{\sf tx}\big]$ and $i \in {\cal S}$. By Lemma~\ref{lem:onion}, we obtain:
\begin{equation}
 \frac{{\sf P}_{\sf tx} |t_i - r_i|^{-\alpha}}{{\sf N_0} + \underset{j \in {\cal S} \backslash \{ i \}}{\sum} {\sf P}_{\sf tx}  |t_j - r_i|^{-\alpha}} \ge
  \frac{{\sf P}_{\sf tx} {\sf r}_{\sf tx}^{-\alpha}}{{\sf N_0} +  {\sf P}_{\sf tx}  {\sf k}(\alpha) ({\sf r}_{\sf xcl} - {\sf r}_{\sf tx})^{-\alpha}}
\end{equation}
Hence,
\begin{equation}
\begin{array}{r@{\ }l}
& {\sf r}_{\sf xcl} \ge
\Big(  \frac{1}{{\sf P}_{\sf tx} {\sf k}(\alpha)} \big( \frac{{\sf P}_{\sf tx}}{\beta}{\sf r}_{\sf tx}^{-\alpha} - {\sf N_0} \big) \Big)^{-\frac{1}{\alpha}} + {\sf r}_{\sf tx} \\
& \Leftrightarrow \
  \frac{{\sf P}_{\sf tx} {\sf r}_{\sf tx}^{-\alpha}}{{\sf N_0} +  {\sf P}_{\sf tx}  {\sf k}(\alpha) ({\sf r}_{\sf xcl} - {\sf r}_{\sf tx})^{-\alpha}} \ge \beta
\ \Rightarrow\
{\cal S} \in {\mathscr U}^{\sf ag}_{\rm sinr}[X,\beta]
\end{array}
\end{equation}
\end{proof}

\begin{lemma}
If ${\sf r}'_{\sf xcl} \ge {\sf r}_{\sf xcl} + 2 {\sf r}_{\sf tx}$, then
\begin{equation}
{\mathscr U}^{\sf pw}_{\rm fr}\big[{X},{\sf r}_{\sf xcl}, {\sf r}_{\sf tx}\big] \supseteq {\mathscr B}^{\sf pw}_{\rm fr}\big[{X},{\sf r}_{\sf xcl}, {\sf r}_{\sf tx}\big] \supseteq {\mathscr U}^{\sf pw}_{\rm fr}\big[{X},{\sf r}'_{\sf xcl}, {\sf r}_{\sf tx}\big]
\end{equation}
\end{lemma}
\begin{proof}
By Lemma~\ref{lem:bi_nf} and set $\Delta = {\sf r}_{\sf xcl} \slash {\sf r}_{\sf tx} -1$ and $\Delta' = {\sf r}'_{\sf xcl} \slash {\sf r}_{\sf tx} -1$
\end{proof}

Lemma~\ref{lem:bi_nf} can be proven by applying Lemma~\ref{lem:bi_pw} and letting ${\sf N_0} = 0$, $\Delta = \beta^{\frac{1}{\alpha}} - 1$ and $\Delta' = \beta'^{\frac{1}{\alpha}} - 1$. For clarity of presentation, we present the proof of Lemma~\ref{lem:bi_nf} in order to reveal the basic idea of Lemma~\ref{lem:bi_pw} in the simpler context when ${\sf N_0} = 0$.

\begin{lemma}
If $\Delta' \ge \Delta + 2$, then
\begin{equation}
{\mathscr U}^{\sf pw}_{\rm sir}\big[X,\Delta\big] \supseteq {\mathscr B}^{\sf pw}_{\rm sir}\big[X,\Delta\big] \supseteq {\mathscr U}^{\sf pw}_{\rm sir}\big[X,\Delta'\big]
\end{equation}
\end{lemma}

\begin{proof}
Follows from Lemma~\ref{lem:bi_ag}.
\end{proof}

\begin{lemma}
If $\beta' \ge ( 2 + \beta^{\frac{1}{\alpha}} )^{\alpha}$, then
\begin{equation}
{\mathscr U}^{\sf pw}_{\rm sinr}\big[X,\beta\big] \supseteq {\mathscr B}^{\sf pw}_{\rm sinr}\big[X,\beta\big] \supseteq {\mathscr U}^{\sf pw}_{\rm sinr}\big[X,\beta'\big]
\end{equation}
\end{lemma}
\begin{proof}
This is proven in a similar fashion as Lemma~\ref{lem:bi_nf}. Suppose ${\cal S} \in {\mathscr B}^{\sf pw}_{\rm sinr}\big[X,\beta\big]$.
First, note that ${\sf P}_{\sf tx} |t_i - r_i|^{-\alpha} \ge \beta {\sf N_0}$ for all $i \in X$. Otherwise, $t_i$ will be unable to transmit to $r_i$ even without interference.
Second,
\begin{equation} \label{eqn:pw_sinr_eq}
\begin{array}{@{}r@{\ }r@{\ }l}
&  \displaystyle \frac{{\sf P}_{\sf tx} |t_i - r_i|^{-\alpha}}{{\sf N_0} +
{\sf P}_{\sf tx} \big({\sf dist}(i,j)\big)^{-\alpha}} & < \beta \\
\Rightarrow &
\Big(  \frac{1}{{\sf P}_{\sf tx}} \big( \frac{{\sf P}_{\sf tx}}{\beta}|t_i - r_i|^{-\alpha} - {\sf N_0} \big) \Big)^{-\frac{1}{\alpha}} & > {\sf dist}(i,j)
\end{array}
\end{equation}
The last inequality is due to the fact that $-\frac{1}{\alpha} < 0$ and  ${\sf P}_{\sf tx} |t_i - r_i|^{-\alpha} \ge \beta {\sf N_0}$.

We need to show that
\begin{equation} \label{eqn:pw_sinr_statement}
{\cal S} \cup \{ i \} \notin {\mathscr B}^{\sf pw}_{\rm sinr}\big[X,\beta\big]
\Rightarrow
{\cal S} \cup \{ i \} \notin {\mathscr U}^{\sf pw}_{\rm sinr}\big[X,( 2 + \beta^{\frac{1}{\alpha}} )^{\alpha}\big]
\end{equation}

Suppose ${\cal S} \cup \{ i \} \notin {\mathscr B}^{\sf pw}_{\rm sinr}\big[X,\beta\big]$ for some given link $i$. Then there are four cases as follows.
\begin{itemize}

\item[$\sf 1$):]
Suppose $\frac{{\sf P}_{\sf tx} |t_i - r_i|^{-\alpha}}{{\sf N_0} +
{\sf P}_{\sf tx}   |t_j - r_i|^{-\alpha}} < \beta$ for some $j \in {\cal S}$. This is trivial that ${\cal S} \cup \{ i \} \notin {\mathscr U}^{\sf pw}_{\rm sinr}\big[X,\beta\big]$.

\item[$\sf 2$):]
Suppose for some $j \in {\cal S}$ that
\begin{equation} \label{eqn:pw_sinr_case2_1}
\frac{{\sf P}_{\sf tx} |t_i - r_i|^{-\alpha}}{{\sf N_0} +
{\sf P}_{\sf tx}   |r_j - r_i|^{-\alpha}} < \beta
\end{equation}
Without loss of generality, we also assume
\begin{equation} \label{eqn:pw_sinr_case2_2}
|t_i - r_i| \ge |t_j - r_j|
\end{equation}
Otherwise, if $|t_j - r_j| \ge |t_i - r_i|$, then
\begin{equation}
\begin{array}{r@{\ }l}
|r_j - r_i| < & \Big(  \frac{1}{{\sf P}_{\sf tx}} \big( \frac{{\sf P}_{\sf tx}}{\beta}  |t_i - r_i|^{-\alpha} - {\sf N_0} \big) \Big)^{-\frac{1}{\alpha}} \\
\le & \Big(  \frac{1}{{\sf P}_{\sf tx}} \big( \frac{{\sf P}_{\sf tx}}{\beta}    |t_j - r_j|^{-\alpha} - {\sf N_0} \big) \Big)^{-\frac{1}{\alpha}}
\end{array}
\end{equation}
Therefore, we can equivalently assume
\begin{equation}
\frac{{\sf P}_{\sf tx}   |t_j - r_j|^{-\alpha}}{{\sf N_0} +
{\sf P}_{\sf tx}   |r_j - r_i|^{-\alpha}} < \beta \mbox{\ and\ } |t_j - r_j| \ge |t_i - r_i|,
\end{equation}
by inter-changing $i$ and $j$.

Next, we obtain:
\begin{equation} \hspace{-20pt}
\begin{array}{@{}r@{}l@{\ }l}
& |t_j - r_i| \le |r_j - r_i| + |t_j - r_j| \\
< & \Big(  \frac{1}{{\sf P}_{\sf tx}} \big( \frac{{\sf P}_{\sf tx}}{\beta}  |t_i \mbox{-} r_i|^{-\alpha} - {\sf N_0} \big) \Big)^{-\frac{1}{\alpha}} + |t_j \mbox{-} r_j| & \mbox{by Eqns.~(\ref{eqn:pw_sinr_case2_1}),(\ref{eqn:pw_sinr_eq})} \\
\le &   \Big(  \frac{1}{{\sf P}_{\sf tx}} \big( \frac{{\sf P}_{\sf tx}}{\beta}  |t_i \mbox{-} r_i|^{-\alpha} - {\sf N_0} \big) \Big)^{-\frac{1}{\alpha}} + |t_i \mbox{-} r_i| & \mbox{by Eqn.~(\ref{eqn:pw_sinr_case2_3})} \\
\le &  (1 + \beta^{-\frac{1}{\alpha}})  \Big(  \frac{1}{{\sf P}_{\sf tx}} \big( \frac{{\sf P}_{\sf tx}}{\beta}  |t_i \mbox{-} r_i|^{-\alpha} - {\sf N_0} \big) \Big)^{-\frac{1}{\alpha}}
\end{array}
\end{equation}
The last inequality is due to the fact that when ${\sf N_0} \ge 0$,
\begin{equation} \label{eqn:pw_sinr_case2_3}
\begin{array}{c}
\frac{\beta}{{\sf P}_{\sf tx}} \big( \frac{{\sf P}_{\sf tx}}{\beta}  |t_i  -  r_i|^{ - \alpha}  -  {\sf N_0} \big) \ge |t_i - r_i|^{-\alpha}
\end{array}
\end{equation}

By Lemma~\ref{lem:ineq1}, setting $\gamma = ( 1 + \beta^{-\frac{1}{\alpha}} )^{\alpha}$, we obtain:
\begin{equation}
\frac{{\sf P}_{\sf tx} |t_i - r_i|^{-\alpha}}{{\sf N_0} +
{\sf P}_{\sf tx}  |t_j - r_i|^{-\alpha}} < ( 1 + \beta^{-\frac{1}{\alpha}} )^{\alpha} \beta = ( 1 + \beta^{\frac{1}{\alpha}} )^{\alpha}
\end{equation}

\item[$\sf 3$):]
Suppose $\frac{{\sf P}_{\sf tx} |t_i - r_i|^{-\alpha}}{{\sf N_0} +
{\sf P}_{\sf tx}   |t_j - t_i|^{-\alpha}} < \beta$ for some $j \in {\cal S}$.  This is shown in a similar way as Case $\sf 2$.

\item[$\sf 4$):]
Suppose for some $j \in {\cal S}$ that
\begin{equation} \label{eqn:pw_sinr_case4_1}
\frac{{\sf P}_{\sf tx} |t_i - r_i|^{-\alpha}}{{\sf N_0} +
{\sf P}_{\sf tx} |t_i - r_j|^{-\alpha}} < \beta
\end{equation}
In addition, we assume
\begin{equation} \label{eqn:pw_sinr_case4_2}
\frac{{\sf P}_{\sf tx}   |t_j - r_j|^{-\alpha}}{{\sf N_0} +
{\sf P}_{\sf tx} |t_i - r_j|^{-\alpha}} \ge \beta
\end{equation}
Otherwise, it reduces to Case $\sf 1$ by inter-changing $i$ and $j$.
Hence, we obtain:
\begin{equation} \label{eqn:pw_sinr_case4_3}
\begin{array}{@{}r@{}r@{\ }l}
& & \Big(  \frac{1}{{\sf P}_{\sf tx}} \big( \frac{{\sf P}_{\sf tx}}{\beta}   |t_j  -  r_j|^{ - \alpha}  -  {\sf N_0} \big) \Big)^{ - \frac{1}{\alpha}}  \\
& \le & |t_i - t_j|
< \Big(  \frac{1}{{\sf P}_{\sf tx}} \big( \frac{{\sf P}_{\sf tx}}{\beta} |t_i  -  r_i|^{ - \alpha}  -  {\sf N_0} \big) \Big)^{ - \frac{1}{\alpha}} \\
\Rightarrow & & |t_j - r_j| < |t_i - r_i|
\end{array}
\end{equation}

Next, we obtain:
\begin{equation} \hspace{-20pt}
\begin{array}{@{}r@{\ }l@{\ }l}
& |t_j - r_i| \le |t_j - r_j| + |r_j - t_i| + |t_i - r_i| \\
< & |t_j  \mbox{-} r_j| + \Big(  \frac{1}{{\sf P}_{\sf tx}} \big( \frac{{\sf P}_{\sf tx}}{\beta}    |t_j \mbox{-} r_j|^{\mbox{-}\alpha} - {\sf N_0} \big) \Big)^{\mbox{-}\frac{1}{\alpha}} + |t_i  \mbox{-} r_i| & \mbox{by Eqn.~(\ref{eqn:pw_sinr_case4_2})} \\
< &  (1 + 2 \beta^{-\frac{1}{\alpha}})  \Big(  \frac{1}{{\sf P}_{\sf tx}} \big( \frac{{\sf P}_{\sf tx}}{\beta}  |t_i \mbox{-} r_i|^{-\alpha} - {\sf N_0} \big) \Big)^{-\frac{1}{\alpha}} & \mbox{by Eqns.~(\ref{eqn:pw_sinr_case4_3}),(\ref{eqn:pw_sinr_case2_3})}
\end{array}
\end{equation}
By Lemma~\ref{lem:ineq1}, setting $\gamma = ( 2 + \beta^{-\frac{1}{\alpha}} )^{\alpha}$, we obtain:
\begin{equation}
\frac{{\sf P}_{\sf tx} |t_i - r_i|^{-\alpha}}{{\sf N_0} +
{\sf P}_{\sf tx}   |t_j - r_i|^{-\alpha}} < \beta  ( 1 + 2 \beta^{-\frac{1}{\alpha}} )^{\alpha} = ( 2 + \beta^{\frac{1}{\alpha}} )^{\alpha}
\end{equation}
\end{itemize}
Therefore, ${\cal S} \cup \{ i \} \notin {\mathscr U}^{\sf pw}_{\rm sinr}\big[X,\beta'\big]$. This proves Eqn.~(\ref{eqn:pw_sinr_statement}).
\end{proof}

\begin{lemma}
If $\beta' \ge  ( 2 + \beta^{\frac{1}{\alpha}} )^{\alpha}$, then
\begin{equation}
{\mathscr U}^{\sf ag}_{\rm sinr}\big[X,\beta\big] \supseteq {\mathscr B}^{\sf ag}_{\rm sinr}\big[X,\beta\big] \supseteq {\mathscr U}^{\sf ag}_{\rm sinr}\big[X,\beta'\big]
\end{equation}
\end{lemma}
\begin{proof}
Suppose ${\cal S} \in {\mathscr B}^{\sf ag}_{\rm sinr}\big[X,\beta\big]$. We need to show
\begin{equation}
{\cal S} \cup \{i\} \notin {\mathscr B}^{\sf ag}_{\rm sinr}\big[X,\beta\big] \ \Rightarrow\
{\cal S} \cup \{i\} \notin {\mathscr U}^{\sf ag}_{\rm sinr}\big[X,( 2 + \beta^{\frac{1}{\alpha}} )^{\alpha}\big]
\end{equation}

First, we assume
\begin{equation} \label{eqn:agg_sinr_1} \hspace{-10pt}
\begin{array}{r@{\ }l}
& \frac{{\sf P}_{\sf tx}  |t_j  -  r_j|^{ - \alpha}}{{\sf N_0}  +
{\sf P}_{\sf tx} \big({\sf dist}(j,i)\big)^{ - \alpha}} \ge \beta \\
\Rightarrow &
\frac{{\sf P}_{\sf tx}  |t_j  -  r_j|^{ - \alpha}}{
{\sf P}_{\sf tx}  \big({\sf dist}(j,i)\big)^{ - \alpha}} \ge \beta
\Rightarrow
|t_j  -  r_j| \le \beta^{ - \frac{1}{\alpha}}  \big({\sf dist}(j,i)\big)
\end{array}
\end{equation}
and
\begin{equation} \label{eqn:agg_sinr_2} \hspace{-10pt}
\begin{array}{r@{\ }l}
& \frac{{\sf P}_{\sf tx}  |t_i  -  r_i|^{ - \alpha}}{{\sf N_0}  +
{\sf P}_{\sf tx}  \big({\sf dist}(i,j)\big)^{ - \alpha}} \ge \beta \\
\Rightarrow &
\frac{{\sf P}_{\sf tx}  |t_i  -  r_i|^{ - \alpha}}{
{\sf P}_{\sf tx}  \big({\sf dist}(i,j)\big)^{ - \alpha}} \ge \beta
\Rightarrow
|t_i  -  r_i| \le \beta^{ - \frac{1}{\alpha}}  \big({\sf dist}(i,j)\big)
\end{array}
\end{equation}
Otherwise, we complete the proof by Lemma~\ref{lem:bi_pw}, such that
\begin{equation}
{\cal S} \cup \{i\} \notin {\mathscr U}^{\sf ag}_{\rm sinr}\big[X, ( 2 + \beta^{\frac{1}{\alpha}} )^{\alpha} \big]
\end{equation}

Next, we obtain:
\begin{equation}
\begin{array}{@{}r@{\ }l@{\ }l}
|t_j - r_i| \le & |t_j - r_j| + |r_j - r_i| \\
\le & \beta^{-\frac{1}{\alpha}}  |r_j - r_i| + |r_j - r_i| & \mbox{by Eqn.~(\ref{eqn:agg_sinr_2})} \\
\Rightarrow |t_j - r_i| \le & (1 + \beta^{-\frac{1}{\alpha}}) |r_j - r_i|
\\ \\
|t_j - r_i| \le & |t_i - r_i| + |t_j - t_i| \\
\Rightarrow  |t_j - r_i| \le & (1 + \beta^{-\frac{1}{\alpha}}) |t_j - t_i| & \mbox{by Eqn.~(\ref{eqn:agg_sinr_1})}
\\ \\
|t_j - r_i| \le & |t_j - r_j| + |r_j - t_i| + |t_i - r_i| \\
\Rightarrow |t_j - r_i| \le & (1 + 2\beta^{-\frac{1}{\alpha}}) |r_j - t_i| & \mbox{by Eqns.~(\ref{eqn:agg_sinr_1}),(\ref{eqn:agg_sinr_2})}
\end{array}
\end{equation}
Therefore,
\begin{equation} \label{eqn:agg_sinr_3}
|t_j - r_i| < (1 + 2\beta^{-\frac{1}{\alpha}}) \cdot {\sf dist}(i,j)
\end{equation}

Also, since  ${\cal S} \cup \{ i \} \notin {\mathscr B}^{\sf ag}_{\rm sinr}\big[X,\beta\big]$,
we obtain:
\begin{equation}
\begin{array}{@{}r@{}r@{}l@{\ }l}
& \frac{{\sf P}_{\sf tx}  |t_i - r_i|^{-\alpha}}{{\sf N_0} +
\underset{j \in {\cal S} \backslash \{ i \}}{\sum} {\sf P}_{\sf tx}  \big({\sf dist}(i,j)\big)^{-\alpha}} & < \beta \\
\Rightarrow & \underset{j \in {\cal S} \backslash \{ i \}}{\sum} \big({\sf dist}(i,j)\big)^{\mbox{-}\alpha}
& > \frac{1}{{\sf P}_{\sf tx}} \big( \frac{{\sf P}_{\sf tx}}{\beta} |t_i \mbox{-} r_i|^{\mbox{-}\alpha} \mbox{-} {\sf N_0} \big) \\
\Rightarrow & \underset{j \in {\cal S} \backslash \{ i \}}{\sum}
 (1 \mbox{+} 2\beta^{\mbox{-}\frac{1}{\alpha}})^{\alpha} |t_j \mbox{-} r_i|^{\mbox{-}\alpha} & >
\frac{1}{{\sf P}_{\sf tx}} \big( \frac{{\sf P}_{\sf tx}}{\beta} |t_i \mbox{-} r_i|^{\mbox{-}\alpha} \mbox{-} {\sf N_0} \big) & \mbox{by Eqn.~(\ref{eqn:agg_sinr_3})}
\end{array}
\end{equation}

By Lemma~\ref{lem:ineq1}, setting $\gamma = ( 2 + \beta^{-\frac{1}{\alpha}} )^{\alpha}$, we obtain:
\begin{equation}
\frac{{\sf P}_{\sf tx}  |t_i - r_i|^{-\alpha}}{{\sf N_0} + \sum_{j \ne i}
{\sf P}_{\sf tx}  |t_j - r_i|^{\alpha}} < \beta (1 + 2\beta^{-\frac{1}{\alpha}})^{\alpha}  = ( 2 + \beta^{\frac{1}{\alpha}} )^{\alpha}
\end{equation}
Finally, we complete the proof by combining with Lemma~\ref{lem:bi_pw}.
\end{proof}

\begin{lemma}
If ${\sf r}_{\sf tx} = \underset{i \in X}{\max} |t_i - r_i|$ and ${\sf r}_{\rm cs} \ge {\sf r}_{\sf xcl} + 2 {\sf r}_{\sf tx}$,
\begin{equation}
{\mathscr C}^{\sf pw}\big[{X},{\sf r}_{\rm xcl}\big]  \supseteq
{\mathscr B}^{\sf pw}_{\rm fr}\big[{X},{\sf r}_{\sf xcl}, {\sf r}_{\sf tx}\big] \supseteq {\mathscr C}^{\sf pw}\big[{X},{\sf r}_{\rm cs}\big]
\end{equation}
\end{lemma}
\begin{proof}
It can be proven in a similar fashion as Lemma~\ref{lem:bi_fr}, where we replace constraint $|t_j - r_i| \ge {\sf r}_{\sf xcl}$ by $|t_j - t_i| \ge {\sf r}_{\rm cs}$.
\end{proof}

\begin{lemma} 
$\langle{\mathscr C}[X],{\bf \nu}\rangle$
is a reversible Markov process. The stationary distribution for each
${\cal S}\in{\mathscr C}[X]$ is: 
\begin{equation} \label{eqn:stationary}
{\mathbb{P}}_{{\bf \nu}}({\cal S})=\frac{\exp\big(\sum_{i\in{\cal S}}\log\nu_{i}\big)}{\sum_{{\cal S}'\in{\mathscr C}[X]}\exp\big(\sum_{j\in{\cal S}'}\log\nu_{j}\big)}
\end{equation}
\end{lemma}


\begin{proof} Eqn.~(\ref{eqn:stationary}) satisfies the detailed
balanced eqn: 
\[
\hspace{-5pt}\begin{array}{@{}r@{}l}
\exp\big(\sum_{j\in\{i\}\cup{\cal S}}\log\nu_{j}\big) & =\exp\big(\sum_{j\in{\cal S}}\log\nu_{j}\big)\cdot\exp\big(\log\nu_{i}\big)\\
\Rightarrow\ {\mathbb{P}}_{{\bf \nu}}(\{i\}\cup{\cal S}) & ={\mathbb{P}}_{{\bf \nu}}({\cal S})\cdot\nu_{i}\end{array}
\]
 Hence, $\langle{\mathscr C}[X],{\bf \nu}\rangle$
is a reversible Markov process, Eqn.~(\ref{eqn:stationary}) is the
stationary distribution. 
\end{proof}

\setcounter{theorem}{3}

\begin{theorem}
There exists a suitable setting of $({\sf r}_{\rm cs}^{\sf B}, {\sf r}_{\rm cs}^{\sf P})$, depending on $\beta$ and the maximum transmission distance in ${X}^{\sf c}$, such that
\begin{equation}
{\mathscr B}^{\sf ag}_{\rm haf}\big[({X}^{\sf B}, \beta), ({X}^{\sf P}, \beta) \big] \supseteq  {\mathscr C}^{\sf pw}_{\rm haf}\big[({X}^{\sf B}, {\sf r}_{\rm cs}^{\sf B}), ({X}^{\sf P}, {\sf r}_{\rm cs}^{\sf P})\big]
\end{equation}
\end{theorem}
\begin{proof}
First, we note that
\begin{equation} \label{eqn:mcs_ag_pw_statement}
\begin{array}{@{}r@{\ }l}
 & {\mathscr B}^{\sf ag}_{\rm haf}\big[({X}^{\sf B}, \beta), ({X}^{\sf P}, \beta) \big] \\
= & \Big\{ \bigcup_{{\sf c} \in \{ {\sf B}, {\sf P}\}} {\cal S}^{\sf c} \mid {\cal S}^{\sf c} \in {\mathscr B}^{\sf ag}_{\rm sinr}\big[{X}^{\sf c},\beta\big] \Big\} \\
& \quad \bigcup \Big\{ {\cal S} \mid i, j \in {\cal S}, \{t_i, r_i \} \cap \{t_j, r_j \} = \varnothing \Big\}
\end{array}
\end{equation}

By Theorem~\ref{thm:hidden}, for each ${\sf c} \in \{{\sf B}, {\sf P} \}$, there exists a suitable ${\sf r}_{\rm cs}^{\sf c}$, depending on $\beta$ and the maximum transmission distance in ${X}^{\sf c}$ such that
\begin{equation}
{\mathscr B}^{\sf pw}_{\rm sinr}\big[{X}^{\sf c},\beta\big] \supseteq {\mathscr C}^{\sf pw}\big[{X}^{\sf c},{\sf r}_{\rm cs}^{\sf c}\big]
\end{equation}
Also, it follows that
\begin{equation} \label{eqn:mcs_ag_pw_cond1}
\begin{array}{@{}c}
\Big\{ \bigcup_{{\sf c} \in \{ {\sf B}, {\sf P}\}} {\cal S}^{\sf c} \mid {\cal S}^{\sf c} \in {\mathscr C}^{\sf pw}\big[{X}^{\sf c},{\sf r}_{\rm cs}^{\sf c}\big] \Big\} \\
\qquad \qquad \qquad \supseteq
{\mathscr C}^{\sf pw}_{\rm haf}\big[({X}^{\sf B}, {\sf r}_{\rm cs}^{\sf B}), ({X}^{\sf P}, {\sf r}_{\rm cs}^{\sf P})\big]
\end{array}
\end{equation}

Next, we need to show there exists another suitable ${\sf r'}_{\rm cs}^{\sf c}$, such that
\begin{equation} \label{eqn:mcs_ag_pw_cond2}
\begin{array}{@{}c}
\Big\{ {\cal S} \mid i, j \in {\cal S}, \{t_i, r_i \} \cap \{t_j, r_j \} = \varnothing \Big\} \\
\qquad \qquad \qquad \supseteq
{\mathscr C}^{\sf pw}_{\rm haf}\big[({X}^{\sf B}, {\sf r'}_{\rm cs}^{\sf B}), ({X}^{\sf P}, {\sf r'}_{\rm cs}^{\sf P})\big]
\end{array}
\end{equation}
If $i \in {X}^{\sf c}$ and $j \in {X}^{\sf c}$, then Eqn.~(\ref{eqn:mcs_ag_pw_cond2}) follows from the tree diagram Fig.~\ref{fig:tree} that there exists a suitable ${\sf r'}_{\rm cs}^{\sf c}$, such that
\begin{equation}
{\mathscr B}^{\sf pw}_{\rm sir}\big[{X}^{\sf c}, {\Delta}^{\sf c}\big] \supseteq {\mathscr C}^{\sf pw}\big[{X}^{\sf c},{\sf r'}_{\rm cs}^{\sf c}\big]
\end{equation}
for any ${\Delta}^{\sf c}>0$.
Else if $i \in {X}^{\sf c}$ and $j \notin {X}^{\sf c}$, then without loss of generality, we consider $i \in {X}^{\sf c}$ and $j \in {X}^{\sf c'}$, and ${\sf r'}_{\rm cs}^{\sf c} > {\sf r'}_{\rm cs}^{\sf c'}$. Then, Eqn.~(\ref{eqn:mcs_ag_pw_cond2}) follows from the fact that there exists a suitable ${\sf r'}_{\rm cs}^{\sf c}$, such that
\begin{equation} \hspace{-5pt}
{\mathscr B}^{\sf pw}_{\rm sir}\big[{X}^{\sf c} \cup \{j\}, {\Delta}^{\sf c}\big] \supseteq {\mathscr C}^{\sf pw}\big[{X}^{\sf c} \cup \{j\},{\sf r'}_{\rm cs}^{\sf c}\big] \supseteq {\mathscr C}^{\sf pw}\big[{X}^{\sf c},{\sf r'}_{\rm cs}^{\sf c}\big]
\end{equation}
for any ${\Delta}^{\sf c}>0$. 

Finally, we take the maximum carrier-sensing range among ${\sf r}_{\rm cs}^{\sf c}$ and ${\sf r'}_{\rm cs}^{\sf c}$, for each ${\sf c} \in \{{\sf P},{\sf B}\}$. Hence, we complete the proof.
\end{proof}

\setcounter{lemma}{8}

\begin{lemma} Given a deterministic scheduling scheme $({\cal S}_{\sf t})_{{\sf t}={\sf1}}^{{\sf m}}$,
let the fraction of time spent in ${\cal S}\in{\mathscr C}[X]$
be ${\mathbb{P}}^{\sf det}({\cal S})=\frac{1}{\sf m}\sum_{{\sf t}=1}^{\sf m}{\mathds1}({\cal S}={\cal S}_{\sf t})$.
If ${\mathbb{P}}^{\sf det}({\cal S})>0$ for all ${\cal S}\in{\mathscr C}[X]$,
then there exists count-down rate ${\bf \nu}$, such that for each
transmitter-receiver pair $i\in X$, it satisfies: \begin{equation}
{\mathfrak{c}}_{i}^{\sf det}\big[({\cal S}_{\sf t})_{{\sf t}={\sf1}}^{{\sf n}}\big]\le{\mathfrak{c}}_{i}^{\sf rand}\big[\langle{\mathscr C}[X],{\bf \nu}\rangle\big]\label{eq:cs_det}\end{equation}
 \end{lemma}

\begin{proof} Let ${\mathfrak{c}}_{i}^{\sf det}\triangleq{\mathfrak{c}}_{i}^{\sf det}\big[({\cal S}_{\sf t})_{{\sf t}={\sf1}}^{{\sf n}}\big]$,
and ${\mathfrak{c}}_{i}^{\sf rand}(\nu)\triangleq{\mathfrak{c}}_{i}^{\sf rand}\big[\langle{\mathscr C}[X],{\bf \nu}\rangle\big]$.
We start by considering the following feasibility problem: \[
\begin{array}{rrl}
\mbox{{\bf P1:}} & \max_{{\bf z}\geq0} & {\displaystyle 0}\\
 & \mbox{subject to} & {\displaystyle \sum_{{\cal S}\in{\mathscr C}[X]:i\in{\cal S}}z_{{\cal S}}\ge{\mathfrak{c}}_{i}^{\sf det},\mbox{for all\ }i\in{X}}\\
 &  & {\displaystyle \sum_{{\cal S}\in{\mathscr C}[X]}z_{{\cal S}}=1}\end{array}\]
Its corresponding Lagranian problem is given by: \[
\begin{array}{rrl}
 & \underset{\mathbf{\lambda}\geq0}{\min}\underset{{\bf z}\geq0}{\max} & -\underset{i\in{X}}{\sum}\lambda_{i}{\mathfrak{c}}_{i}^{\sf det}+\underset{{\cal S}\in{\mathscr C}[X]}{\sum}z_{{\cal S}}\sum_{i\in{\cal S}}\lambda_{i}\\
 & \mbox{subject to} & {\displaystyle \sum_{{\cal S}\in{\mathscr C}[X]}z_{{\cal S}}=1}\end{array}\]
 The subproblem in $\mathbf{z}$ is the Maximum Weighted Independent
Set (MWIS) problem, which is a combinatorial problem, whose optimal
value is given on the left hand side of the following formula and
is approximated by the log-sum-exp function on the right hand side
\cite{boyd2004convex}:

\begin{equation}
\underset{{\cal S}\in{\mathscr C}[X]}{\max}\sum_{i\in{\cal S}}\lambda_{i}\approx\log\Big(\sum_{{\cal S}\in{\mathscr C}[X]}\exp\big(\sum_{i\in{\cal S}}\lambda_{i}\big)\Big).\label{eq:approx}\end{equation}

Define the log-sum-exp function as $g\left(\boldsymbol{\lambda}\right)$.
It is known that ts conjugate function is a convex function and is
given by \cite{boyd2004convex} \begin{equation}
g^{*}\left(\boldsymbol{z}\right)=\begin{cases}
\sum_{{\cal S}\in{\mathscr C}[X]}z_{{\cal S}}\log z_{{\cal S}} & \mbox{if }\boldsymbol{z}\geq0\mbox{ and }\boldsymbol{1}^{T}\boldsymbol{z}=1\\
\infty & \mbox{otherwise.}\end{cases}\label{eq:thm.1.1}\end{equation}
Further, $g\left(\boldsymbol{\lambda}\right)$ is convex and closed;
hence, the conjugate of its conjugate $g^{*}\left(\boldsymbol{z}\right)$
is itself. Specifically, $g\left(\boldsymbol{\lambda}\right)$ is
the optimal value of a concave optimization problem: \begin{eqnarray}
g\left(\boldsymbol{\lambda}\right) & = & \underset{{\bf z}\geq0}{\max}\underset{{\cal S}\in{\mathscr C}[X]}{\sum}z_{{\cal S}}\sum_{i\in{\cal S}}\lambda_{i}{\displaystyle -\sum_{{\cal S}\in{\mathscr C}[X]}z_{{\cal S}}\log(z_{{\cal S}})}\label{eq:thm.1.2}\\ 
 &  & \mbox{subject to}{\displaystyle \sum_{{\cal S}\in{\mathscr C}[X]}z_{{\cal S}}=1}\nonumber \end{eqnarray}
Therefore, by the log-sum-exp approximation in Eq. (\ref{eq:approx}),
we implicitly solve an approximated version of the MWIS problem, off
by an \emph{entropy }term $-\sum_{{\cal S}\in{\mathscr C}[X]}z_{{\cal S}}\log(z_{{\cal S}})$.
By solving the Karush-Kuhn-Tucker (KKT) conditions \cite{boyd2004convex},
we obtain the optimal solution to the approximated problem in Eq.
(\ref{eq:thm.1.2}) as follows, \begin{equation}
z_{{\cal S}}\left(\boldsymbol{\lambda}\right)=\frac{\exp\big(\sum_{i\in{\cal S}}\lambda_{i}\big)}{\sum_{{\cal S}'\in{\mathscr C}[X]}\exp\big(\sum_{j\in{\cal S}'}\lambda_{j}\big)}.\label{eq:sol.MA.MWIS.appox}\end{equation}
Comparing Eq. (\ref{eqn:stationary}) and (\ref{eq:sol.MA.MWIS.appox}),
we observe that $z_{{\cal S}}\left(\boldsymbol{\lambda}\right)$ is
the percentage of time state $\mathcal{S}$ being activated under
CSMA scheduling, with every link $i\in X$ counts down with a rate
of $\nu_{i}=\exp\left(\lambda_{i}\right)$.

After the log-sum-exp approximation in Eq. (\ref{eq:approx}), the
Lagranian problem becomes

\begin{equation}
\begin{array}{rrl}
 & \underset{\mathbf{\lambda}\geq0}{\min}\underset{{\bf z}>0}{\max} & -\underset{i\in{X}}{\sum}\lambda_{i}{\mathfrak{c}}_{i}^{\sf det}+\underset{{\cal S}\in{\mathscr C}[X]}{\sum}z_{{\cal S}}\sum_{i\in{\cal S}}\lambda_{i}\\
 &  & \;\;-\sum_{{\cal S}\in{\mathscr C}[X]}z_{{\cal S}}\log(z_{{\cal S}})\\
 & \mbox{subject to} & {\displaystyle \sum_{{\cal S}\in{\mathscr C}[X]}z_{{\cal S}}=1}\end{array}\label{lag.prob:p2}\end{equation}
We reverse-engineer to get the corresponding primal problem as the
following entropy maximization problem: \[
\begin{array}{rrl}
\mbox{{\bf P2:}} & \max_{{\bf z}\geq0} & {\displaystyle -\sum_{{\cal S}\in{\mathscr C}[X]}z_{{\cal S}}\log(z_{{\cal S}})}\\
 & \mbox{subject to} & {\displaystyle \sum_{{\cal S}\in{\mathscr C}[X]:i\in{\cal S}}z_{{\cal S}}\ge{\mathfrak{c}}_{i}^{\sf det},\mbox{for all\ }i\in{X}}\\
 &  & {\displaystyle \sum_{{\cal S}\in{\mathscr C}[X]}z_{{\cal S}}=1}\end{array}\]
By the setting in the lemma statement, the linear constraint set of
problem \textbf{P2} contains at least one relative interior point,
and thus problem \textbf{P2} satisfies Slater's condition \cite{boyd2004convex}.
Consequently, the optimal dual and primal solutions to its Lagranian
problem in Eq. (\ref{lag.prob:p2}) $\boldsymbol{\lambda}^{*}$ and
$\boldsymbol{z}_{{\cal S}}\left(\boldsymbol{\lambda}^{*}\right)$
exist, and $\boldsymbol{z}_{{\cal S}}\left(\boldsymbol{\lambda}^{*}\right)$
is the optimal solution to problem \textbf{P2} and is feasible. In
other words, by CSMA scheduling with every link $i\in X$ counts down
with a rate of $\nu_{i}=\exp\left(\lambda_{i}^{*}\right)$, we obtain
that for every link $i\in X$ its link throughput satisfies: \begin{equation}
{\mathfrak{c}}_{i}^{\sf rand}(\nu)=\sum_{{\cal S}\in{\mathscr C}[X]:i\in{\cal S}}\boldsymbol{z}_{{\cal S}}\left(\boldsymbol{\lambda}^{*}\right)\ge{\mathfrak{c}}_{i}^{\sf det}.\end{equation}
This completes the proof.
\end{proof}

\setcounter{lemma}{13}
\begin{lemma} \label{lem:cs_pw_pw}
Let ${\sf r}_{\sf tx} = {\max}_{i \in X} |t_i - r_i|$. If
\begin{equation}
{\sf r}_{\sf xcl} \ge
\Big(  \frac{1}{{\sf P}_{\sf tx}} \big( \frac{{\sf P}_{\sf tx}}{\beta}{\sf r}_{\sf tx}^{-\alpha} - {\sf N_0} \big) \Big)^{-\frac{1}{\alpha}} 
\end{equation}
then
\begin{equation}
{\mathscr U}^{\sf pw}_{\rm sinr}[X,\beta] \supseteq  {\mathscr U}^{\sf pw}_{\rm fr}\big[{X},{\sf r}_{\sf xcl}, {\sf r}_{\sf tx}\big]
\end{equation}
\end{lemma}
\begin{proof}
Suppose ${\cal S} \in {\mathscr U}^{\sf pw}_{\rm fr}\big[{X},{\sf r}_{\sf xcl}, {\sf r}_{\sf tx}\big]$ and $i,j \in {\cal S}$. Then,
\begin{equation}
\begin{array}{r@{\ }l}
& {\sf r}_{\sf xcl} \ge
\Big(  \frac{1}{{\sf P}_{\sf tx}} \big( \frac{{\sf P}_{\sf tx}}{\beta}{\sf r}_{\sf tx}^{-\alpha} - {\sf N_0} \big) \Big)^{-\frac{1}{\alpha}}  \\
& \Leftrightarrow \
  \frac{{\sf P}_{\sf tx} {\sf r}_{\sf tx}^{-\alpha}}{{\sf N_0} +  {\sf P}_{\sf tx}  {\sf r}_{\sf xcl}^{-\alpha}} \ge \beta
  \Leftrightarrow \
  \frac{{\sf P}_{\sf tx} |t_i - r_i|^{-\alpha}}{{\sf N_0} +  {\sf P}_{\sf tx} |t_j - r_i|^{-\alpha}} \ge \beta
\end{array}
\end{equation}
for all $i, j \in {\cal S}$. Hence, ${\cal S} \in {\mathscr U}^{\sf ag}_{\rm sinr}[X,\beta].$
\end{proof}

\begin{lemma} 
Let ${\sf r}_{\sf tx} = {\max}_{i \in X} |t_i - r_i|$. If
\begin{equation}
{\sf r}_{\rm cs} \ge
\Big(  \frac{1}{{\sf P}_{\sf tx}} \big( \frac{{\sf P}_{\sf tx}}{\beta}{\sf r}_{\sf tx}^{-\alpha} - {\sf N_0} \big) \Big)^{-\frac{1}{\alpha}} + 2{\sf r}_{\sf tx}
\end{equation}
then
\begin{equation}
{\mathscr U}^{\sf pw}_{\rm sinr}[X,\beta] \supseteq  {\mathscr C}^{\sf pw}\big[{X},{\sf r}_{\sf cs}\big]
\end{equation}
\end{lemma}
\begin{proof}
Follows from Lemma~\ref{lem:cs_pw_pw}, \ref{lem:bi_pw}, \ref{lem:cs_fr}.
\end{proof}

\end{document}